\documentclass[aps,amsfonts,preprint]{revtex4-2}
\usepackage[utf8]{inputenc}
\usepackage[T1]{fontenc}
\usepackage{tgtermes}
\usepackage{amsmath, amssymb}
\usepackage{mathtools}
\usepackage{breqn}
\usepackage{graphicx}
\usepackage{float}
\usepackage{bm}
\usepackage{physics}
\usepackage{xcolor}
\usepackage{verbatim}
\usepackage{enumitem}
\usepackage{chemformula}
\usepackage{hyperref}
\usepackage{multirow,xcolor}
\usepackage{chemformula}
\usepackage{makecell}
\hypersetup{
    colorlinks = true,
    linkcolor = blue,
    linkbordercolor = white,
    citecolor=blue,
    urlcolor=black
}
\usepackage[export]{adjustbox}
\makeatletter
\let\cat@comma@active\@empty
\makeatother

\begin{abstract}
    Electron-electron scattering is one of the most important hot carrier relaxation pathways in plasmonic nanoparticles.  Understanding the dynamics of this scattering process and the effects of this on excited state dephasing and relaxation is therefore essential for the design of plasmonic nanostructures, including optical properties and the dynamics of electrons in plasmon-driven catalytic reactions.
    In this work, we have developed a novel approach that incorporates real-time time-dependent density functional tight-binding (DFTB) simulations with the Lindblad quantum Boltzmann equation (LQBE) based on a screened electron-electron interaction that is determined by the random phase approximation (RPA). This approach enables a self-consistent description of electron-electron scattering effects that occur during and after plasmon excitation in clusters/nanoparticles with hundreds of atoms.
    With our RT-TDDFTB+LQBE method, we investigate the quasiparticle lifetime as well as population and coherence dynamics in silver, gold and aluminum nanoclusters with sizes between 1.5-2.6 nm.
    Our results show that the quasiparticle lifetimes and relaxation dynamics are highly energy dependent, becoming much faster at higher energies. For clusters less than 2 nm, quantum effects associated with discrete energy levels can lead to the fluctuating lifetimes and deviation of the dynamics from the typical thermalization process, while for larger nanoparticles, the transition to bulk metallic behavior is found. Decoherence of the initially excited plasmon resonance is observed with a timescale of 10 fs, much faster than population relaxation.
    For gold, we find that the 5d-band can significantly slow down the relaxation of energetic electrons though Auger scattering, and interband transitions can lead to a secondary decoherence process longer than 50 fs.
    Our method provides a general framework for incorporating electron-electron scattering in the dynamics of metallic systems and demonstrates the capability of tight-binding methods to accurately describe electron dynamics for timescales up to ps for clusters/nanoparticles that transition from molecular to metallic properties.
\end{abstract}

\begin{document}

\title{Real-Time Electron-Electron Scattering Dynamics in Plasmonic Nanostructures}
\author{Yanze Wu}
\email{wuyanze@northwestern.edu}
\affiliation{Department of Chemistry, Northwestern University, Evanston, Illinois 60208, USA}
\author{George C. Schatz}
\email{g-schatz@northwestern.edu}
\affiliation{Department of Chemistry, Northwestern University, Evanston, Illinois 60208, USA}

\maketitle
\newpage

Excitation of localized surface plasmons in nanostructures offers a clean and tunable way to enhance photochemistry, opening new avenues in photocatalysis and solar energy harvesting.\cite{zhang2018SurfacePlasmonDriven} Hot carriers (HC) that are produced by plasmon excitation are believed to be key intermediate species in plasmon induced catalysis,\cite{zhang2018SurfacePlasmonDriven, zhang2019PlasmonDriven,kazuma2019Mechanistic} so the lifetimes and relaxation properties of these carriers are crucial.
Studies have shown that HC relaxation typically starts within 100 fs after illumination,\cite{linic2015Photochemical,luo2023Plasmoninduced} while theoretical modeling shows that plasmon driven reactions can happen on a similar timescale.\cite{dallosto2024Peeking,giri2023Photodissociation,herring2023Mechanistic,zhang2018Plasmonic} Therefore, an accurate description of HC relaxation in the modeling of plasmon driven reactions is important to understanding the enhancement mechanism.

Common electron relaxation pathways in plasmonic systems include electron-electron (e-e), electron-phonon (e-ph) scattering and spontaneous radiation.\cite{besteiro2017Understanding} For relatively small nanoparticles, the fastest energy damping channel is e-e scattering, which has been reported to be on the order of 100 fs-1 ps.\cite{vanzan2024Theoretical,pettine2023EnergyResolved,zhao2024Relaxation,wach2025dynamics} However, the most common approach to plasmonic catalysis modeling -- the time-dependent density functional theory (TDDFT), is not able to to capture e-e scattering effects due to the adiabatic approximation used.
Theoretically, e-e scattering in nanostructures has been modeled by phenomenological damping (also known as ``thermalization time'' in modeling spectroscopic properties),\cite{besteiro2017Understanding,voisin2004Ultrafast,manjavacas2014PlasmonInduced} as incorporated into rate equations \cite{zhang2021Theory,saavedra2016HotElectron} and the quantum Boltzmann equation (QBE) with a self-consistent scattering rate \cite{liu2018Relaxation} or from bulk lifetimes.\cite{brown2016Nonradiative,brown2017Experimental,seibel2023TimeResolved} DFT with an empirical damping has also been reported.\cite{govorov2015Kinetic,joao2023Atomistic} Many of the approaches rely on jellium models or empirical damping, which limits their compatibility with TDDFT and transferability between cluster types and compositions. Therefore, we feel it necessary to develop a new extension to TDDFT that incorporates the e-e scattering in a non-empirical, self-consistent way.

As a commonly used formalism of TDDFT, real-time TDDFT (RT-TDDFT) \cite{li2020RealTime,xu2024RealTime} has been applied extensively to the atomistic modeling of plasmonic systems.\cite{herring2023Recent} RT-TDDFT has the advantage of describing the light pulse in real time and is also computationally fast for large systems. As the electronic density matrix is propagated explicitly, RT-TDDFT is also friendly for extensions (e.g., adding external field or dissipation effects).
In recent years, the time-dependent density functional tight-binding (DFTB) method\cite{elstner1998Selfconsistentcharge} has been successfully applied to the modeling of plasmonic physics and chemistry.\cite{chellam2025Density,cuny2018Densityfunctional, giri2023Photodissociation,douglas-gallardo2019Plasmoninduced, dagostino2018Density} By using a minimal basis and a factorized Coulomb repulsion integral, DFTB significantly reduces computational cost, making it applicable to large systems (hundreds to thousands of atoms) while still maintaining a reasonable description of quantum effects. Despite being a semi-empirical method, the self-consistent charge (SCC-)DFTB has a similar Hamiltonian structure with DFT (in fact, several hybrid approaches proposed recently have almost filled the gap between DFT and DFTB \cite{asadi-aghbolaghi2020TDDFT+TB,giannone2020Minimal,grimme2016Ultrafast}), therefore most theories built on the top of TDDFT can be migrated to TDDFTB as well.

In this work, we extend the RT-TDDFT(B) method with a Lindblad quantum Boltzmann equation that incorporates e-e scattering effects. The damping parameters are evaluated self-consistently from the random-phase approximation (RPA) screened potential during the propagation. A minimal non-Markovian kernel is applied to reduce artifacts associated with the incorrect treatment of initial correlations.
Using our new approach, we investigate the electronic lifetime and relaxation time in icosahedral gold, silver and aluminum clusters with different sizes. We find that the orbital lifetimes in our nanoclusters are similar to the bulk values, while the aggregated relaxation times fluctuate with particle sizes.
We also find that the relaxation dynamics is highly heterogeneous, affected by the hot carrier energy and the band structure.

\section{Results and Discussions}

\subsection{Theory of Real-Time TDDFT and Electron-Electron Scattering} \label{sec:theory}

In RT-TDDFT, one propagates the one-body density matrix according to the quantum Liouville equation:
\begin{align}
    \dv{\rho_{pp'}}{t} = -\frac{i}{\hbar}[h^{\text{DFT}}(t),\rho(t)]_{pp'} \label{eq:rttddft}
\end{align}
where $\rho$ and $h^{\text{DFT}}$ are the density matrix and the time-dependent Hamiltonian in the spatial molecular orbital (MO) basis (given by a ground-state calculation), respectively. The elements of $h^{\text{DFT}}$ are given by
\begin{align}
    h^{\text{DFT}}_{pp'}(t) = h^0_{pp'}(t) + \sum_{qq'}{v_{pp',qq'}^{H}\rho_{qq'}(t)} + v^{xc}_{pp'}[\rho,t] \label{eq:h_dft}
\end{align}
In Eq.~\eqref{eq:h_dft}, $h^0_{pp'}(t)$ is the one-body Hamiltonian, including the electron-nuclear interaction and external field terms, and $v^{H}_{pp',qq'}$ and $v^{xc}_{pp'}$ are the Coulomb operator and exchange-correlation (XC) potential, respectively.
 
According to the Runge-Gross theorem \cite{runge1984Densityfunctional,gross1985Local}, the exact $v^{xc}$ depends on not only the current density $\bm{\rho}(t)$, but also its entire evolution history. In practice, RT-TDDFT simulations often apply the adiabatic approximation, where $v^{xc}$ depends only on the instantaneous density (typically with a functional designed for the ground state). As pointed by Maitra et al,\cite{maitra2004Double} TDDFT without a frequency-dependent kernel cannot describe doubly excited states well, and e-e scattering obviously involves double excitations. Though auger-like scattering events have been reported in RT-TDDFT/ALDA simulations for metals, however this scattering is much slower and has incorrect energy scaling.\cite{kononov2022Electron} 
Therefore, an extension beyond the adiabatic TDDFT is necessary to incorporate e-e scattering.

A widely used technique for modeling e-e scattering is the quantum Boltzmann equation (QBE). In the spatial MO basis, QBE reads (note that in some other context, QBE is defined also with transport terms, but here we only include the collision integral)
\begin{align}
    \dv{f_p}{t} &= -2\sum_{qrs}{{\Gamma}_{pqrs}f_p(1-f_q)f_r(1-f_s)} \nonumber\\
    &\qquad+ 2\sum_{qrs}{{\Gamma}_{qpsr}f_q(1-f_p)f_s(1-f_r)} \label{eq:qbe}
\end{align}
where $f_p = \rho_{pp}/2$ is the orbital occupation, and the prefactor 2 accounts for scattering with both same and different spin electrons. The rate coefficient $\Gamma$ is given by Fermi's Golden rule:\cite{echenique2000Theory}
\begin{align}
    \Gamma_{pqrs} = \frac{2\pi}{\hbar}\abs{W_{pq,rs}(\omega_{p}-\omega_q)}^2\delta(\omega_p-\omega_q+\omega_r-\omega_s) \label{eq:Gamma}
\end{align}
where $W_{pq,rs}(\omega)$ is the frequency-dependent screened Coulomb potential (here we neglect the higher order exchange, which is a common practice in metal simulations) and $\omega$'s are the orbital energies. In this work, $W$ is evaluated according to the random-phase approximation (RPA) theory, which is known to be work well in metals.\cite{echenique2000Theory}
For systems with finite sizes, the delta function needs to be replaced by an analytical function.  (In this work, we use gaussian functions. See Sec.~S2C in the Supporting Information (SI) for details.)

The QBE has been widely applied in modeling e-e scattering in metals and nanostructures.\cite{rethfeld2002Ultrafast,liu2018Relaxation,brown2017Experimental,seibel2023TimeResolved} The standard QBE only considers the orbital populations, and therefore has to be extended to the Lindbladian form to describe the evolution of the density matrix. In the simplest extension (which has been referred as the ``dephasing-rate approximation'' \cite{haas1996Generalized}), the off-diagonal density matrix components $\rho_{pp'}$ are damped according to the sum of the inflow and outflow rate coefficients, and averaged between the two orbitals $p$ and $p'$ \cite{binder1992Carriercarrier,bohne1990Phase,kadanoff2000Quantum} (For a detailed explanation, see Sec.~S1A in the SI). We then arrive at the Lindblad-QBE (LQBE):
\begin{align}
    \dv{\rho_{pp'}}{t} = \mathcal{L}_{\text{QBE}}(\rho)_{pp'} =& -\sum_{qrs}{({\Gamma}_{pqrs}+{\Gamma}_{p'qrs})\rho_{pp'}(1-f_q)f_r(1-f_s)} \nonumber\\
    &+ \sum_{qrs}{({\Gamma}_{qpsr}+\Gamma_{qp'sr})f_s(1-f_r)f_q(2\delta_{pp'}-\rho_{pp'})} \label{eq:lqbe}
\end{align}
Note that the diagonal collision terms of the Eq.~\eqref{eq:lqbe} are the same as the standard QBE. By combining LQBE with RT-TDDFT, we have a Markovian equation:
\begin{align}
    \dv{\rho_{pp'}}{t} = -\frac{i}{\hbar}[h^{\text{DFT}}(t),\rho(t)]_{pp'} + \mathcal{L}_{\text{QBE}}(\rho)_{pp'} \label{eq:lqbe_markov}
\end{align}

While Eq.~\eqref{eq:lqbe_markov} can already capture e-e scattering effects, it has some difficulties in treating the short time dynamics during laser excitation. The reason is that the Markovian QBE (or LQBE) assumes electrons (or holes) to be well-defined quasiparticles which always evolve at their orbital energies. Such assumptions may not hold during the initial stage of the laser pulse where some short-lived ``off-resonant'' transitions exist between source and destination orbitals with energy gap different from the laser energy (see Sec.~S1B and S2A in the SI for a short discussion; also see Section 6.4 of Ref.~\cite{bonitz2016Quantum}). In this case, the Markovian LQBE often leads to excessive absorption of the laser pulse energy and violation of total energy conservation (see Fig.~S4 in the SI). While an ultimate solution would be a full non-Markovian kernel with all frequency components of the density matrix included, this would be computationally prohibitive for large systems. Instead, in this work, we apply a minimal non-Markovian correction to the QBE, which only filters out the density matrix components that do not evolve according to the orbital energies (see Sec.~S1B in the SI):
\begin{align}
    \bar{\rho}_{pp'}(t) &= \begin{dcases} \gamma\int_0^t{\rho_{pp'}(t-\tau)e^{i\omega_{pp'}\tau-\gamma \tau}d\tau } & p\ne p' \\
    \rho_{pp}(t) - \sum_{q\ne p}{\frac{\abs{\rho_{pq}(t)}^2 - \abs{\bar{\rho}_{pq}(t)}^2}{\rho_{pp}(0)-\rho_{qq}(0)} } & p=p'
    \end{dcases}
\end{align}
By replacing $\rho$ by $\bar{\rho}$ in Eq.~\eqref{eq:lqbe_markov}, we arrive at the final equation of motion:
\begin{align}
    \dv{\rho_{pp'}}{t} = -\frac{i}{\hbar}[h^{\text{DFT}}(t),\rho(t)]_{pp'} + \mathcal{L}_{\text{QBE}}(\bar{\rho})_{pp'} \label{eq:lqbel_nonmarkov}
\end{align}
In our tests, the non-Markovian correction has very little effect on the population distribution, while greatly improving energy absorption and conservation at short time (see Sec.~S2A in the SI). As a side effect, we have excluded the e-e scattering from the off-resonant transitions. Such processes are related to the resistive absorption \cite{brown2016Nonradiative} (also referred as electron-assisted-absorption \cite{khurgin2019Hot}) of the metal.
Nevertheless, for small nanoparticles, the Drude effects are relatively weak,\cite{brown2016Nonradiative} and the plasmonic mode is also short-lived due to strong decoherence effects (typically around 10 fs, see the coherence subsection), therefore we do not expect any significant impact of resistive absorption on the moderate to long time dynamics (10-300 fs), which is the main focus of this paper.  

In the literature, QBE calculations have been reported directly using Kohn-Sham (KS) orbital energies or density-of-states (DOS).\cite{mueller2013Relaxation, brown2017Experimental} However, since the QBE is usually formalized from many-body perturbation theory (MBPT), it is worth to give a few more words on its compatibility with DFT.  
It has been known that the exact exchange-correlation (XC) functional can be connected with MBPT.\cite{casida2012Progress} While the exact relation is involved, a simpler understanding starts from the quasiparticle GW framework where the real part of the self-energy ($\Sigma$) has a similar role to the XC potential:\cite{faleev2004AllElectron, kotani2007Quasiparticlea,golze2019GW} 
\begin{align}
    v^{xc}_{pq} \approx \frac{1}{2}(\Re[\Sigma_{pq}(\omega_p)]+\Re[\Sigma_{pq}(\omega_q)])
\end{align}
The on-diagonal difference, $v^{xc}_{pp}-\Re[\Sigma_{pp}]$ is known as the quasiparticle (QP) corrections. For metals, the QP corrections are typically very small (< 0.2 eV for orbitals within 2 eV of the Fermi surface),\cite{cazzaniga2012approaches} therefore we expect that $\Re[\Sigma]$ and $v_{xc}$ can be used interchangeably with reasonable accuracy. The imaginary part of the self-energy, however, has no counterpart in adiabatic TDDFT.
For metals, to the lowest order, the imaginary self-energy can be approximated by the Born approximation, which in the end gives the exact same equation as the LQBE (see Ref.~\cite{kadanoff2000Quantum}, also Sec.~S1A in the SI). 
Admittedly, from the Kramers-Kronig relation, the imaginary self-energy will be always accompanied by a frequency dependent real-valued potential, which corresponds to including some memory effects and initial correlations.\cite{bonitz1996Numerical} Nevertheless, these effects are shown to be weak in metals where the screening is strong.\cite{bonitz1996Numerical} Due to the substantial computational cost of including these self-energy effects, we choose to keep a simple description where only the QBE terms are included.

\subsection{Model Systems and Lifetime Profile} \label{sec:simulation}

\begin{figure}[h]
    \centering
    \includegraphics[width=0.7\linewidth]{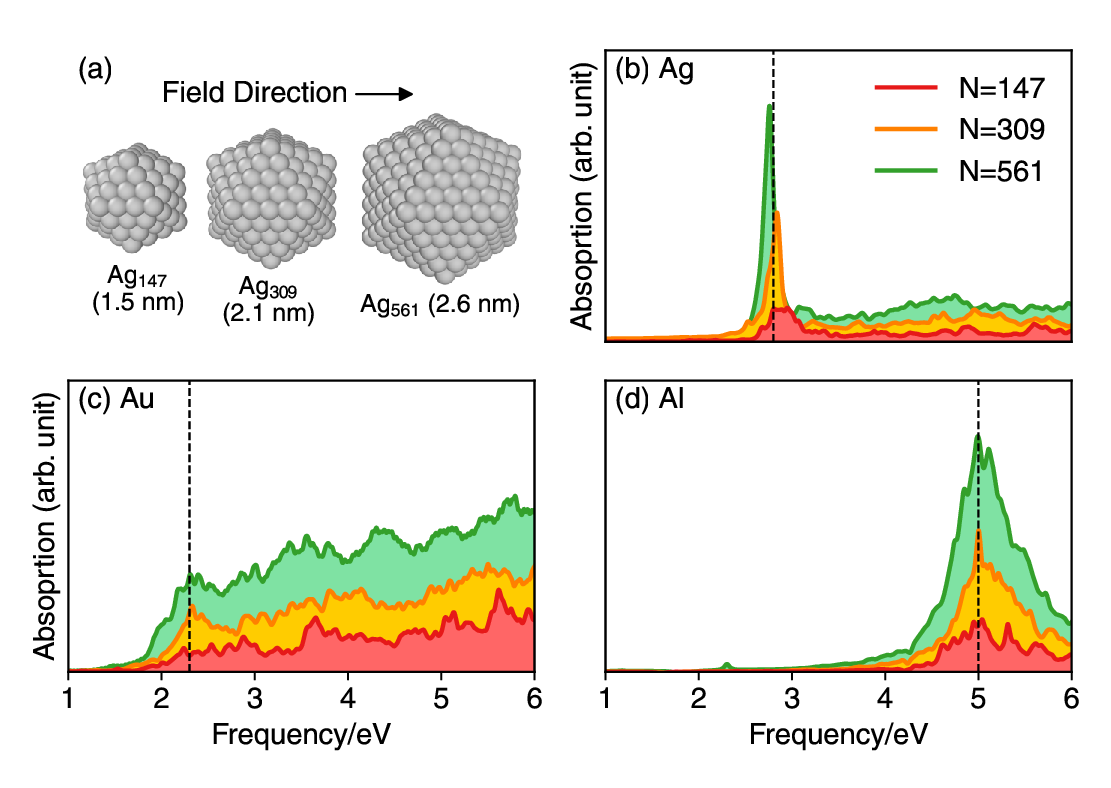}
    \caption{(a) Silver clusters used in the simulations. The gold and aluminum clusters are similar in size. (b)-(d): The absorption spectra of silver, gold and aluminum clusters, respectively, calculated by the standard RT-TDDFTB simulation with delta-kick perturbation. The approximate location of plasmon resonance frequency is marked by the dashed line.}
    \label{fig:spec}
\end{figure}

We focus on the neutral icosahedral clusters (\ch{X_{147}}, \ch{X_{309}} and \ch{X_{561}}) with X=Ag, Au and Al. 
The clusters with different elements have similar sizes: 1.5 nm for \ch{X_{147}}, 2.1 nm for \ch{X_{309}} and 2.6 nm for \ch{X_{561}}, see Fig.~\ref{fig:spec}a. The initial electronic structures are minimized with an effective electron temperature of 300 K within the framework of self-consistent-charge (SCC)-DFTB. To identify the plasmon frequency, we perform standard RT-TDDFTB simulation with delta-kick perturbation \cite{yabana1996Timedependent} to evaluate the absorption spectra (See the Method Section). As shown in Fig.~\ref{fig:spec}b-d, the spectra of silver and aluminum show clear plasmonic peaks around 2.8 eV and 5 eV, respectively, while the plasmon peak of gold at 2.3 eV is overlapped with strong absorption from interband transitions.

As a test of the applicability of the QBE with DFTB, we evaluate the quasiparticle (QP) lifetime in all clusters. The QP lifetime determines the average free time for a carrier before colliding with another electron or hole. In Green's function theory, where the QPs are defined in terms of resonances of the single-body Green's function at equilibrium, the lifetime of a QP state $p$ is given by the imaginary part of its self-energy according to \cite{echenique2000Theory} 
\begin{align}
    \frac{1}{\tau_p}=-\frac{2}{\hbar}\Im[\Sigma_{pp}(\omega_p)]   \label{eq:gw_lifetime}
\end{align}
In the QBE framework, the QP lifetime is given by:\cite{1991Landau}
\begin{align}
    \frac{1}{\tau_p^{\text{QBE}}} = 2\sum_{qrs}{(\Gamma_{pqrs}(1-f_q^{\text{eq}})f_r^{\text{eq}}(1-f_s^{\text{eq}}) + \Gamma_{qpsr}f_q^{\text{eq}}f_r^{\text{eq}}(1-f_s^{\text{eq}}))} \label{eq:qbe_lifetime}
\end{align}
where $f^{\text{eq}}$ is the equilibrium population (here we use the Fermi Dirac distribution at 300 K).

Below we will compare lifetimes from this formula for Ag, Au and Al clusters with the results of several other calculations as well as experimental measurements.
For the theoretical values, we include the lifetime from Fermi liquid theory (FLT), where the lifetime obeys a simple quadratic relation with the orbital energy:\cite{echenique2000Theory, qian2005Lifetime} 
\begin{align}
    \frac{1}{\tau_{\text{FLT}}(E)} = Ar_s^{-5/2}((E-E_F)^2+(\pi k_BT)^2)
\end{align}
where $A=262.6\text{ fs}^{-1}\text{eV}^{-2}$ is the prefactor and $r_s$ is the effective density parameter. For silver and gold, we have scaled $r_s$ according to Ref.~\cite{bauer2015Hot} and \cite{pettine2023EnergyResolved} to account for screening from d-band, which is not explicitly present in FLT. We also include several GW calculations in the bulk phase \cite{zhukov2001Corrected, ladstadter2004Firstprinciples} where the QP lifetimes are evaluated according to Eq.~\eqref{eq:gw_lifetime}. For experimental data, we include the two-photon photoemission (2PPE) measurements for gold and silver nanoparticles \cite{merschdorf2004Collective,pettine2023EnergyResolved} as well as for bulk aluminum.\cite{bauer1998Electron}

\begin{figure}[H]
    \centering
    \includegraphics[width=\linewidth]{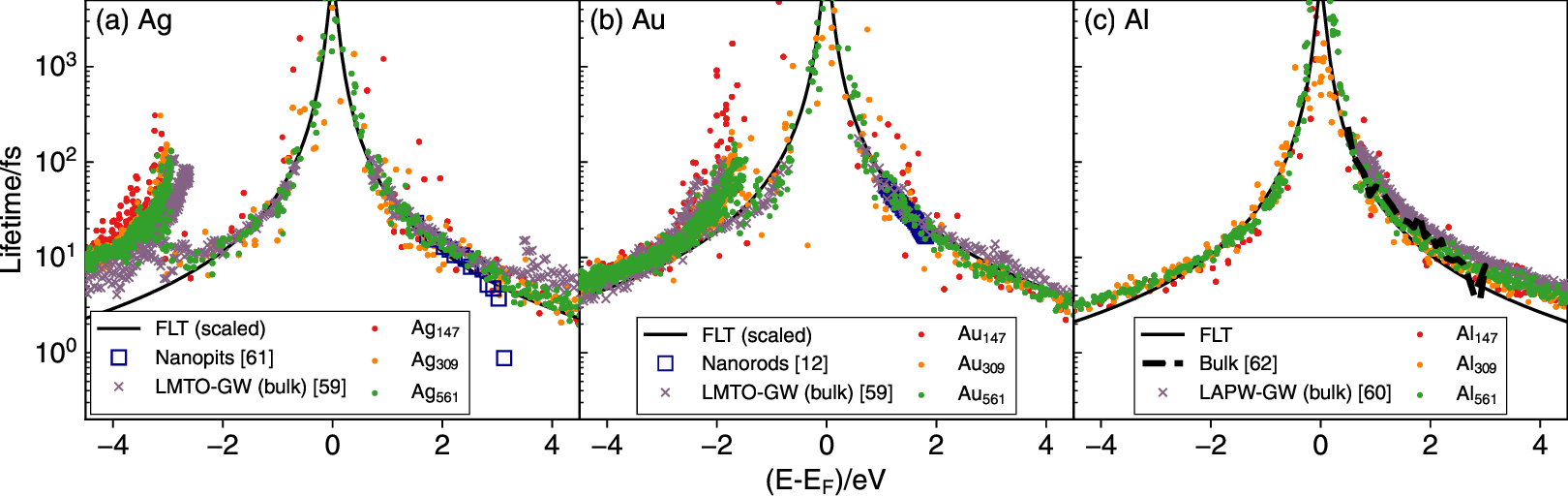}
    \caption{QBE lifetime calculated according to Eq.~\eqref{eq:qbe_lifetime}, compared with various existing experimental and theoretical results. (a) For silver, we compare with 2PPE measurements of nanopits (20 nm width, 2 nm height) on graphite \cite{merschdorf2004Collective}, and a linear Muffin-tin orbital (LMTO) GW calculation for the bulk.\cite{zhukov2001Corrected} (b) For gold, we compare with 2PPE measurements of nanorods (40 nm length, 10 nm diameter) \cite{pettine2023EnergyResolved} and the LMTO-GW calculation for the bulk.\cite{zhukov2001Corrected} (c) For aluminum, we compare with 2PPE measurements of the bulk phase \cite{bauer1998Electron} and a linearized-augmented-plane-wave (LAPW) GW calculation for the bulk.\cite{ladstadter2004Firstprinciples} The free electron density parameters used are $r_s[\text{Ag}]=2.0, r_s[\text{Au}]=1.87$ and $r_s[\text{Al}]=2.07$.\cite{bauer2015Hot,pettine2023EnergyResolved}}
    \label{fig:lifetime}
\end{figure} 

As shown in Fig.~\ref{fig:lifetime}, despite the various differences in experimental conditions and theoretical frameworks, the lifetimes provided by DFTB-QBE shows overall good agreement with the existing results for all the elements, and capture the characteristic $(E-E_F)^{-2}$ decay at higher energies as well as the long lifetime of the d-band in noble metals. Considering the small size (around 2 nm) of our systems, this implies that the QP lifetime is not greatly affected by the size of the nanostructure {\em per se} via any internal screening effects, in contrary to certain models.\cite{voisin2004Ultrafast} Note that in practice, there are still many other factors that can lead to a size dependence in the measured lifetime, such as transport effects and the screening of the surrounding media.\cite{bauer1998Electron} 
The lifetimes reported here are also shorter than a previous estimation from jellium model with a semiclassical screening function, which suggests the importance of a correct treatment of screened potential.\cite{saavedra2016HotElectron}

For gold, the d-band hole lifetime at the band edge ($\abs{E-E_F}\approx\text{-1.6 eV}$) is 40-100 fs, significantly longer than the value of sp-band (around 20-50 fs at the same energy) and the FLT prediction, consistent with previous modeling and experimental studies.\cite{lee2023Band,zhukov2001Corrected,campillo2000Hole} 
Such a long lifetime has a strong impact in the electron relaxation dynamics in gold, as we will show in the following sections.
Note that a much shorter d-band lifetime is observed in recent studies due to the frequent e-ph scattering,\cite{bernardi2015Theory} however we expect these e-ph scatterings to have a minor effect on the population dynamics on the 100 fs scale,\cite{lee2023Band} as the energy change per scattering is relatively small.

For small clusters (especially when N=147), the lifetime shows larger fluctuations near the Fermi surface due to the more discrete spectrum of orbital energies. The net result of the fluctuations is often an increase of the lifetime (though exceptions are also common), due to the lack of relaxation channels. Such fluctuations may lead to a deviation of the overall relaxation rate, together with other kinetic effects, as we will show below. 
Following this trend, we expect that for tiny clusters with N < 100, certain orbitals may not find suitable relaxation channels and populations there may be trapped for a considerable time. Such trapping mechanisms have also been reported in electron-phonon relaxation studies of tiny gold clusters.\cite{zhou2017Evolution,varnavski2010Critical}

\subsection{Real-Time Dynamics} \label{sec:pop}

The RT-TDDFTB dynamics is initiated with a gaussian laser pulse polarized along the $x$ direction, with the field strength $E=E_0e^{-4\pi t^2/\sigma^2}\sin(\omega_{\text{exc}}t)$, where $E_0=0.01$ V/Angstrom (1.33$\times 10^9 \text{ W/cm}^2$) and $\sigma=\text{10 fs}$. The laser frequency is $\omega_{\text{exc}}$ chosen to be same as the localized surface plasmonic resonance frequency $\omega_{\text{LSPR}}$ of each cluster (see the Method Section for details). 

\begin{figure}[H]
    \centering
    \includegraphics[width=0.7\linewidth]{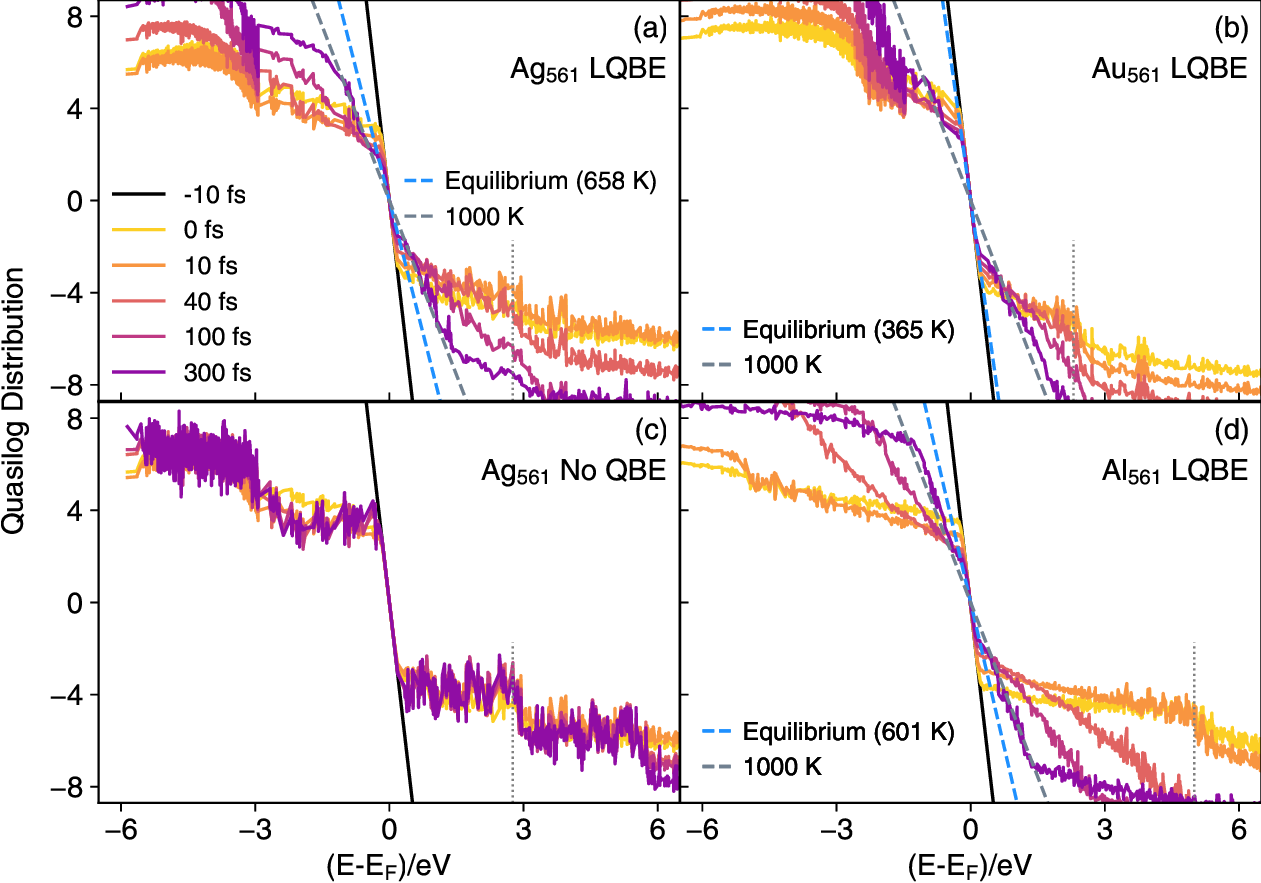}
    \caption{The quasilogarithmic distribution $\phi(E)=-\log_{10}{(1/f(E)-1)}$ for RT-TDDFTB+LQBE simulations of (a) \ch{Ag_{561}} (b) \ch{Au_{561}} and (d) \ch{Al_{561}}, and (c) the standard RT-TDDFTB simulation of \ch{Ag_{561}}. The laser frequency is marked by the vertical dotted lines. The distributions at  1000 K and the terminal temperature are shown by dashed lines. In all systems, RT-TDDFTB+LQBE shows population relaxation as a result of e-e scattering, while in the standard RT-TDDFTB simulation, the distribution has no significant change after the laser pulse up to 300 fs. }
    \label{fig:dist}
\end{figure}

To characterize the population dynamics in e-e scattering processes in nanoclusters, we plot the quasi-logarithmic (quasilog) occupation distribution $\phi(E)=-\log_{10}{(1/f(E)-1)}$ in Fig.~\ref{fig:dist}(a,b,d) at several simulation times and 300K. We also plot the thermal distribution at both 1000 K and at the equilibrium temperature after long simulation (different for each system). 
For comparison, the quasilog distribution from a standard RT-TDDFTB simulation (i.e., without QBE terms) for \ch{Ag_{561}} is plotted in Fig.~\ref{fig:dist}c. 
Before the excitations, the quasilog distribution is a straight line corresponding to the Fermi-Dirac distribution at 300 K. Initially after the excitation, the occupation forms a high temperature non-equilibrium distribution, with Fermi steps appearing at multiples of the excitation frequency. When the e-e scattering is taken into account, the Fermi steps are quickly washed away, as also found in previous studies.\cite{lugovskoy1999Ultrafast,rethfeld2002Ultrafast}

In RT-TDDFTB+LQBE simulations, the slope in the quasilog distribution becomes steeper over time, indicating cooling down of the system. At 300 fs, the distributions for all clusters reaches an effective temperature around 1000 K, but are still far from the thermal equilibrium.
The quasilog distribution in the standard RT-TDDFTB simulation (i.e., without e-e scattering terms) shows no relaxation at all (Fig.~\ref{fig:dist}c).

\begin{figure}[H]
    \centering
    \includegraphics[width=0.75\linewidth]{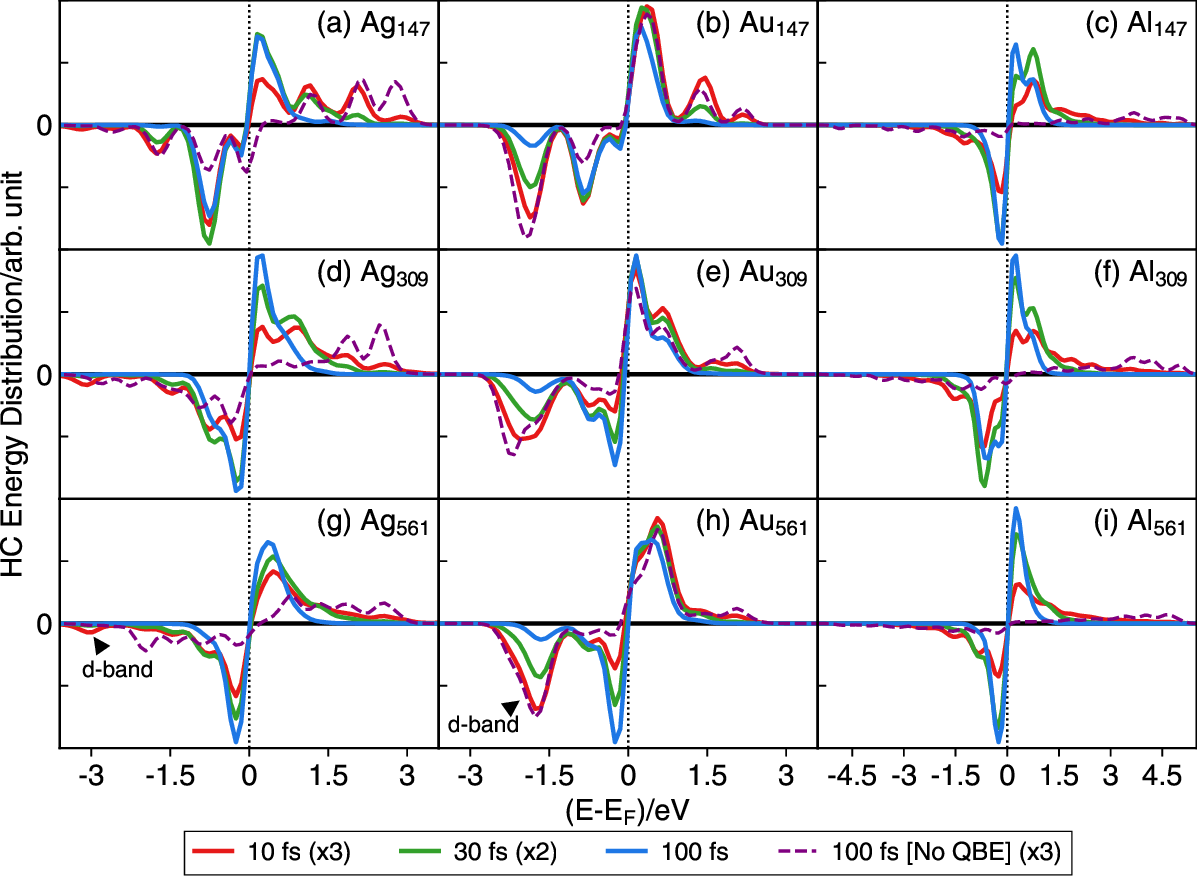}
    \caption{Energy distribution of hot carriers among different clusters at 10 fs, 30 fs and 100 fs after the laser pulse peak, along with the distribution in simulations without QBE terms at 100 fs. Note that we have scaled some of the distributions to make better comparison. The actual change is largely monotonic with time. }
    \label{fig:dist_smeared_linear}
\end{figure}

Another characterization of relaxation process is the energy distribution of HCs, defined as
\begin{align}
    \Delta P(E) = (f(E,t)-f(E,0))g(E)
\end{align}
with $g(E)$ being the DOS. Previous theoretical studies have shown that the hot carriers generated by plasmon decay already concentrate near the Fermi surface within 100 fs or so after the laser pulse,\cite{zhang2021Theory, brown2017Experimental, voisin2004Ultrafast,saavedra2016HotElectron} though a longer time is reported in Ref.~\cite{liu2018Relaxation}. With our new method, we are able to quantify the change of energy distribution within the first 100 fs including the full set of QBE scattering rates.
In Fig.~\ref{fig:dist_smeared_linear}, we plot the energy distribution of hot carriers in our simulation with LQBE at 10, 30 and 100 fs after the peak of the laser pulse, along with the distribution in simulations without e-e scattering terms at 100 fs. A negative distribution indicates holes and a positive distribution indicates electrons. 

Initially after the laser excitation, the hot carriers exhibit a fairly broad distribution across energies up to the excitation energy (about 2.8 eV for Ag, 2.3 eV for Au and 5.0 eV for Al). 
The distribution of smaller clusters shows more individual, small peaks due to the sparsity of energy levels, which is also shown in the TDDFT study in Ref.~\cite{rossi2020HotCarrier}. The silver and gold clusters show d-band peaks around -3 and -1.7 eV from the Fermi surface, respectively, as also reported in previous modeling studies.\cite{sundararaman2014Theoretical,bernardi2015Theory,joao2023Atomistic} For gold clusters, the hot electrons excited from interband transitions also result in strong peaks at around 0.5 eV. In our calculations, the plasmonic resonance frequency is below the d-band of silver, therefore the d-band absorption is relatively weak and transient compared to previous modeling studies.\cite{rossi2020HotCarrier,sundararaman2014Theoretical,bernardi2015Theory,joao2023Atomistic}

As the relaxation proceeds, the population gradually evolves towards the Fermi surface. At 100 fs after the laser pulse, the sp-band HC beyond 1.5 eV is almost completely relaxed. The 100 fs distribution of sp-band carriers shows no significant difference between all clusters, except in certain small clusters (\ch{Ag_{147}},\ch{Au_{147}}) where the holes are peaked slightly further from the Fermi surface due to the distribution of DOS (see Fig.~S11 in the SI). 
For gold clusters, due to the large DOS and longer lifetime (see Fig.~\ref{fig:lifetime}b), the d-band holes in gold are still evident at 100 fs. The prolonged existence of d-band holes suggests they may be chemically active, which aligns with the hot-hole mediated catalytic mechanism described in earlier work.\cite{schlather2017Hot,ahlawat2021Plasmoninduced,lyu2023Photocatalysis} 
In simulations without e-e scattering, there is still a high population of HCs at high energies at 100 fs.
Overall, our population data largely confirms the fast relaxation of the distribution observed in previous studies, and also show the significance of the d-band in the long time distributions. Compare to the standard RT-TDDFT(B) simulation, our method can capture change of distribution at long times, which allows for a more detailed description of slow dynamics (including chemical processes).
%

\subsection{Relaxation Time} \label{sec:dynamics}

\begin{figure}[h]
    \centering
    \includegraphics[width=0.95\linewidth]{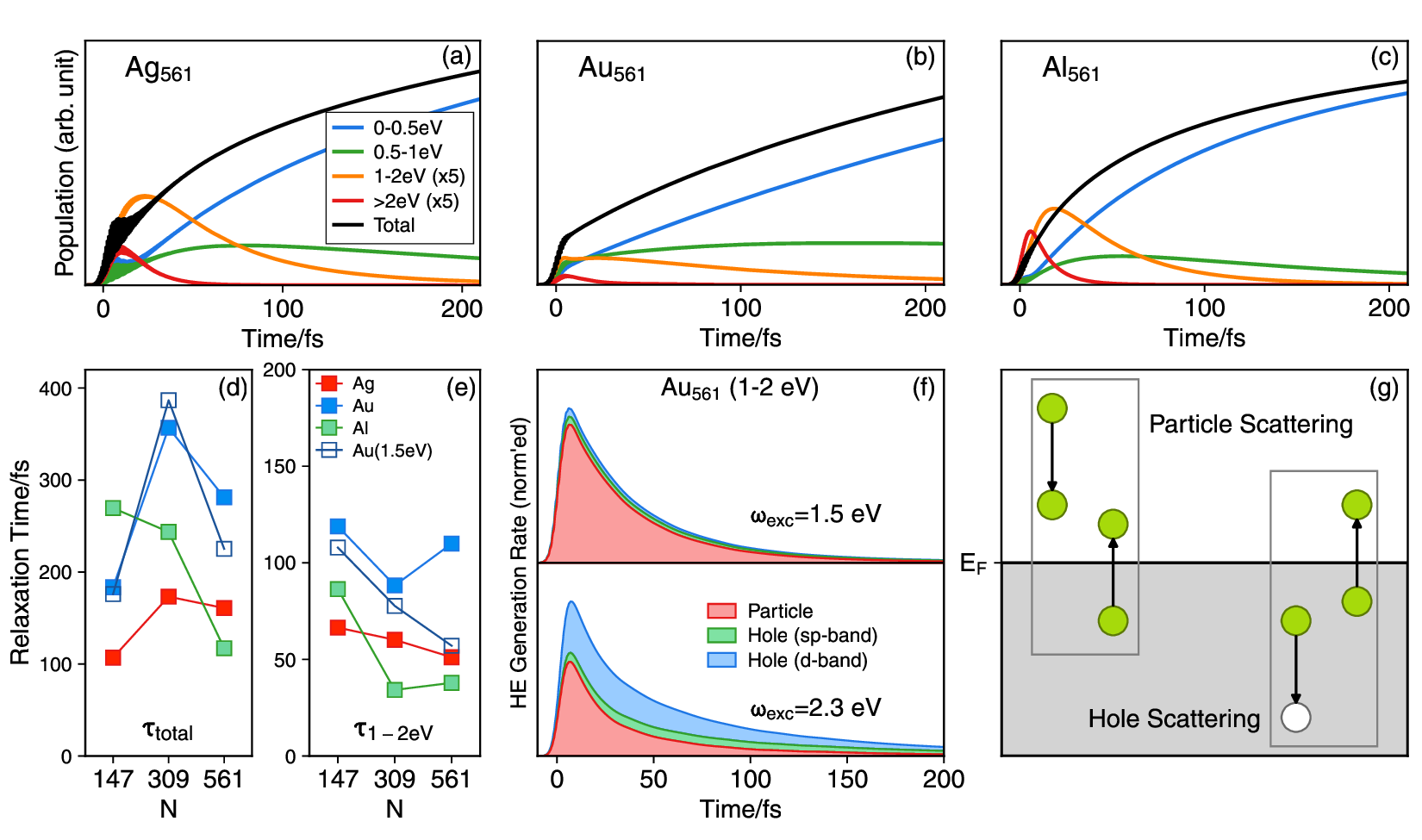}
    \caption{(a-c) HE population in (a) \ch{Ag_{561}} (b) \ch{Au_{561}} and (c) \ch{Al_{561}} as a function time, grouped by energy relative to the Fermi surface. The high energy orbitals relax significantly faster than low energy ones.
    (d)(e) The effective HE relaxation time of clusters, calculated by exponentially fitting (d) the total HE population or (e) the 1-2 eV population. (f) The HE generation rate in \ch{Au_{561}} at different excitation energies according to Eq.~\eqref{eq:scattering_rate}, with the destination energy $E_D$ between 1-2 eV relative to $E_F$. For particle scattering, the source energy $E_S>0$, for sp-band hole scattering $\text{-1.6 eV}<E_S<0$, and for d-band hole scattering $E_S<\text{-1.6 eV}$. For $\hbar\omega_{\text{exc}}$=2.3 eV, hole scattering from the d-band is a major source of HE generation. (g) Schematic view of particle and hole scattering. In the hole scattering, a newly produced electron fills an existing hole (marked by white circle). 
    }
    \label{fig:pop}
\end{figure}

The relaxation time of hot carriers after plasmon excitation is an important factor in understanding the mechanism of plasmon driven reactions. Due to the heterogeneity of particle lifetime, the population dynamics is highly energy dependent. In Fig.~\ref{fig:pop}a-c, we plot the hot electron (HE) population grouped by energy from the Fermi surface in our RT-TDDFTB+LQBE simulations. The HE population >1 eV shows exponential decay over time, while for lower energies (<0.5 eV), the HE population keeps increasing from both the relaxation of more energetic species and from excitation of electron-hole pairs due to scattering. 

In modeling and experimental studies of plasmonic nanostructures, however, e-e relaxation is often characterized by a single thermalization time. Experimentally reported thermalization times vary from 100-700 fs for gold nanoclusters \cite{voisin2004Ultrafast, zhao2024Relaxation, wach2025dynamics, brown2017Experimental, link1999Spectral} and 150-200 fs for silver.\cite{voisin2004Ultrafast,bigot2000Electron} In this work, to investigate the relaxation time with different energy species, we consider the following two definitions of the relaxation time:
\begin{enumerate}
    \item \textit{The growth of total HC} (either HE or hot holes (HH), since they are equal) \textit{population}. This is closer to the classical definition of ``thermalization time'', and can be connected to transient absorption spectroscopy (TAS) measurements from a lower band or the core shell. Since we do not include e-ph coupling, the total HC population increases monotonically over time, therefore we fit it by $P_{\text{total}}=A - Be^{-t/\tau_{\text{total}}}$. The results are shown in Fig.~\ref{fig:pop}d.
    \item \textit{The dissipation of HE that are 1-2 eV above the Fermi surface.} This represents the lifetime of the chemically active species, and can be associated with a 2PPE experiment. After the laser excitation, the population 1-2 eV above the Fermi surface will first increase and then exponentially decay. In the decay stage, the population can be fitted by $P_{\text{1-2eV}} = Ae^{-t/\tau_{\text{1-2eV}}}$. The results are shown in Fig.~\ref{fig:pop}e.
\end{enumerate}
To investigate the effect of d-band (the band edge is $E_F-\text{1.6 eV}$ in our calculations), we also include the simulation results of gold clusters with an excitation frequency of 1.5 eV.

For all clusters, $\tau_{\text{total}}$ falls in the range 100-500 fs that is typical of TAS results. Meanwhile $\tau_{\text{1-2eV}}$ is around 50-100 fs, much shorter than $\tau_{\text{total}}$, which is consistent with the shorter lifetime of high energy electrons. The gold clusters generally show longer lifetimes than silver and aluminum due to longer individual QP lifetimes (see Fig.~\ref{fig:lifetime}, and also Fig.~S13 in the SI), which agrees with the longer relaxation time reported in experiments.
As the cluster sizes are generally small, the relaxation times (especially $\tau_{\text{total}}$) often show large fluctuations between clusters with different sizes due to various quantum and kinetic effects. Specifically, in our systems, we have identified the following three causes of relaxation time variation: 

First: fluctuations of the QP lifetimes. For example, in \ch{Au_{309}}, the long lifetime of several orbitals near 1 eV and -1 eV leads to a substantial increase of $\tau_{\text{total}}$. In general, smaller clusters lack relaxation channels and tend to have a longer relaxation time, which leads to an overall decreasing trend in the relaxation times (especially $\tau_{\text{1-2eV}}$) with increasing cluster size. 

Second: kinetic effects from the distribution of DOS. In general, when the cluster size is reduced, the orbital energies become more discretized and the dynamics gradually deviates from the typical bulk thermalization picture in various ways. For example, in \ch{Ag_{147}} and \ch{Au_{147}}, due to the low DOS around 0.5-1 eV, most of the HEs > 1 eV directly relax to 0-0.5 eV without first going to an intermediate species, therefore effectively reducing the relaxation time $\tau_{\text{total}}$ (see Fig.~S14 in the SI).

Third: d-band hole scattering in gold clusters. In general, the e-e scattering in metals can be divided into two types: particle scattering (relaxation of a hot electron) and hole scattering (relaxation of a hot hole, also referred as Auger scattering in the literature), as shown in Fig.~\ref{fig:pop}g. Due to the high d-band energy (-1.6 eV in our calculation) and high hole state density, hole scattering can continuously generate Auger electrons above 1 eV, which can strongly impact hot electron relaxation. To quantify the contribution of d-band scattering, we further evaluate the energy-specific HE generation rate:
\begin{align}
    r(E_S\to E_D) = \begin{dcases}
        \sum_{p,q,r,s}{(\delta(\hbar\omega_p-E_S)(\delta(\hbar\omega_q-E_D)+\delta(\hbar\omega_s-E_D))}&\\
        \qquad\times\Gamma_{pqrs}f_p(1-f_q)f_r(1-f_s)) & E_S > E_F \\
        \sum_{p,q,r,s}{\delta(\hbar\omega_q-E_S)\delta(\hbar\omega_s-E_D)\Gamma_{pqrs}f_p(1-f_q)f_r(1-f_s)} & E_S < E_F \end{dcases} \label{eq:scattering_rate}
\end{align}
The two cases in Eq.~\eqref{eq:scattering_rate} correspond to particle scattering and hole scattering shown in Fig.~\ref{fig:pop}g, respectively.

As shown in Fig.~\ref{fig:pop}f for \ch{Au_{561}}, when $\hbar\omega_{\text{exc}}=\text{2.3 eV}$, a significant amount of HEs within 1-2 eV are generated from the d-band hole scattering, becoming even more significant than the particle scattering at longer time. By switching the excitation frequency from the $\omega_{\text{LSPR}}$ (about 2.3 eV) to 1.5 eV, the d-band hole scattering is eliminated, and the $\tau_{\text{1-2eV}}$ is also reduced (see Fig.~\ref{fig:pop}e). Such d-band induced longer relaxation times were also observed in a recent experiment.\cite{zhao2024Relaxation} 
For \ch{Au_{147}} and \ch{Au_{309}}, the d-band scattering has less effect on $\tau_{\text{1-2eV}}$, because the electron lifetime in the 1-2 eV range is already long and comparable to that for the d-band holes. The $\tau_{\text{total}}$ is also less affected by the d-band scattering for all clusters, as the majority of HEs are distributed close to the Fermi surface and have a much longer lifetime than the d-band holes.

Overall, we show that the relaxation dynamics can be quite diverse in small clusters and the relaxation time may have a strong system dependence. For moderate-to-large gold clusters, the d-band hole scattering could be a key source of high energy hot electrons, which may have an important role in chemical processes.

\subsection{Coherence Dynamics} \label{sec:coherence}

In the standard picture of plasmon relaxation, the collective plasmonic excitation is transformed into individual electron-hole excitations in the first 10s of fs, a process known as Landau damping or surface scattering.\cite{yannouleas1992Landau} The physical picture is that, due to the surface effects, the plasmon mode interacts with dark states (mostly the individual electron-hole excitations) and they form superposition states. After being excited by the light, these superposition states will gradually lose coherence due to differences in excited state energies, and the off-diagonal density matrix will decay exponentially (known as ``dephasing'') .\cite{yannouleas1992Landau,bhasin2025Plasmon} 
A characteristic of decoherence is the exponential decay of the the dipole moment. For small clusters, the dark states may have big energy spacings, and the dynamics may show recurrence before decoherence is completed. This has been confirmed by dynamical simulations, where the dipole moment shows recurrences after initial decay.\cite{rossi2020HotCarrier,habib2023Machine,bhasin2025Plasmon}
In the frequency domain, the spectra of these small clusters will split into several individual peaks (sometimes referred as ``Landau fragmentation'').\cite{reinhard1996sum, yannouleas1989Fragmentation,lerme2010Size,aikens2008Discrete} 

\begin{figure}[h]
    \centering
    \includegraphics[width=0.9\linewidth]{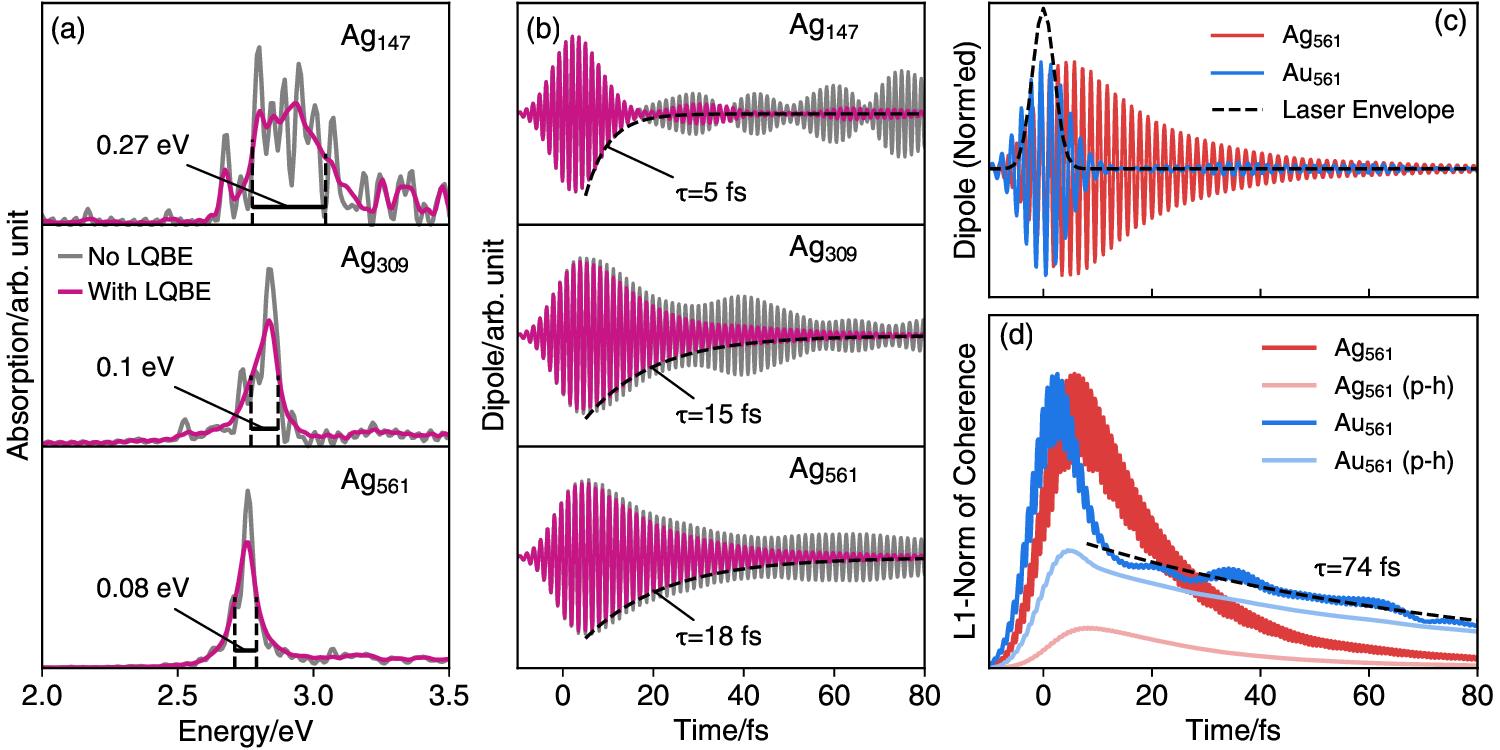}
    \caption{(a) The absorption cross section of silver nanoclusters, calculated by the delta-kick perturbation method, modeled by standard RT-TDDFTB (gray curves) and RT-TDDFTB+LQBE (magenta curves). The spectra of standard RT-TDDFTB are more fragmented in smaller clusters. The horizontal bar and number marks the full-width-half-maximum (FWHM) of the spectra.
    (b) The $x$-direction dipole moment component as a function of time for silver clusters, in dynamics initiated with a finite width laser pulse, modeled by standard RT-TDDFTB (gray curves) and RT-TDDFTB+LQBE (magenta curves). The number marks the lifetime evaluated from the exponential fit of the envelope. In all the standard RT-TDDFTB simulations, the dipole shows recurrence after the initial decay. (c) The $x$-direction dipole moment component from RT-TDDFTB+LQBE simulations for \ch{Ag_{561}} and \ch{Au_{561}}. The envelope of the laser is also shown in black dashed curve. (d) The normalized $C_{l_1}$ and $C_{l_1}^{\text{p-h}}$ in RT-TDDFTB+LQBE simulations for \ch{Ag_{561}} and \ch{Au_{561}}. The second decay stage of $C_{l_1}$ of \ch{Au_{561}} is exponentially fitted and marked by the black dashed curve.
    }
    \label{fig:deco}
\end{figure}

In theoretical modeling, if the further decay of hot carriers is not taken into account (e.g., in adiabatic TDDFT), the absorption spectrum will be a series of discrete peaks even for relatively large clusters.\cite{bhasin2025Plasmon} In practice, an artificial broadening is often required in the post processing to connect with experiment. With our LQBE extension, these peaks are naturally smoothed out. Here we use silver clusters as the representative system as their plasmonic peaks are more distinct. The data for gold and aluminum clusters are included in Fig.~S16 in the SI.
In Fig.~\ref{fig:deco}a, we plot the spectra of three silver clusters with different sizes, modeled by the delta-kick perturbation method with both standard RT-TDDFTB and RT-TDDFTB+LQBE (See the Method Section). In the standard RT-TDDFTB simulations, the spectra are highly fragmented in smaller clusters (especially \ch{Ag_{147}}), while simulations with LQBE provide relatively smooth spectra. This indicates that the fragmentation is greatly suppressed by e-e scattering.

Similar observations are found in the time domain. In Fig.~\ref{fig:deco}b, we plot the $x$-direction component of the dipole moment as a function of time in simulations with a finite pulse. In the standard RT-TDDFTB simulation, the dipoles for all clusters show recurrence. The recurrence appears to be stronger than reported in previous RT-TDDFT simulations,\cite{rossi2020HotCarrier} which is related to our use of an unoptimized geometry (which reduces the randomness) and possibly also to the approximations in DFTB.
With the LQBE terms, the recurrence is effectively prohibited by the damping of off-diagonal elements, therefore the dipole amplitudes generally show monotonic decay (though a slight recurrence is still seen in \ch{Ag_{147}}). This indicates that for clusters as small as 150 atoms, the intrinsic fragmentation and recurrence are already weak due to the e-e scattering. 

By exponentially fitting the envelope of the dipole moment in Fig.~\ref{fig:deco}b, we obtain a dephasing time around 10-15 fs for silver clusters, which is consistent with the widths of spectra in Fig.~\ref{fig:deco}a. This value is longer than the experimental value of 5-10 fs,\cite{luo2023Plasmoninduced,hovel1993Width} but similar to previous TDDFT simulations after removing the artificial broadening.\cite{aikens2008Discrete,kuisma2015Localized} The dephasing time would be further reduced if the Drude friction and e-ph coupling of plasmonic carriers are taken into account.\cite{brown2016Nonradiative} As shown by the time-versus-dipole plot in Fig.~\ref{fig:deco}c, the dipole moment in gold almost decays to zero shortly after the laser period, which is much faster than silver.

Although dipole moment is commonly used as the measurement of coherence, in certain cases it is not accurate enough, as the coherence may exist between ``dark states'' which contribute little to the dipole operator. As an alternative measurement, we evaluate the L1-norm of coherence,\cite{baumgratz2014Quantifying} which is the sum of the absolute value of all off-diagonal density matrix elements:
\begin{align}
    C_{l_1} = \sum_{p\ne q}{\abs{\rho_{pq}}} 
\end{align}
We also define ``coherence of particle-hole pairs'' as the sum of the $\abs{\rho_{pq}}$ where the energy difference between orbital $p,q$ is within $\hbar\Delta\omega=\text{0.5 eV}$ of the excitation energy:
\begin{align}
    C_{l_1}^{\text{p-h}} = \sum_{p\ne q}{\abs{\rho_{pq}}}, \text{  where }\omega_{\text{exc}}-\Delta\omega < \abs{\omega_p-\omega_q} < \omega_{\text{exc}}+\Delta\omega
\end{align}
In the standard dephasing picture, at longer times, the electronic wavefunction will become close to individual particle-hole (p-h) excitations, therefore $C_{l_1}^{\text{p-h}}$ should eventually dominate the coherence (see Fig.~S17 in the SI, and also Ref.~\cite{rossi2020HotCarrier}). With our LQBE extension, the HCs can experience further relaxation, therefore $C_{l_1}^{\text{p-h}}$ will also slowly decay.

As shown in Fig.~\ref{fig:deco}d, the coherence in silver mainly comes from the collective states (instead of individual p-h excitations), therefore the $C_{l_1}$ shows a monotonic decay, matching the dipole dynamics. Due to the short lifetime of sp-band carriers, the $C_{l_1}^{\text{p-h}}$ also decays within 60 fs.
For gold, the coherence of collective states is relatively weak and decays quickly during the laser period, and a second decoherence stage arises at longer time from the particle-hole transitions through e-e scattering. The majority of the particle-hole pairs in gold are interband transitions (See Fig.~S17 in the SI).
Due to the long lifetime of d-band carriers, the decoherence time in the second stage is 74 fs from exponential fit, much longer than the typical dephasing time of plasmonic resonances. 
The existence of two decoherence stages in gold indicates that the coherence dynamics can be complicated when multiple bands are involved. The strong interband coherence in gold nanoclusters is also reported in a recent modeling study.\cite{berdakin2020Interplay} 
Note that for bulk (or a large cluster), the decoherence of interband transitions can be accelerated by the strong e-ph coupling according to Ref.~\cite{bernardi2015Theory}, therefore the slow decoherence stage may not be observable, except at very low temperature, when e-ph relaxation is added to the theory. For small nanoclusters, the effect of e-ph scattering on the interband coherence remains to be explored. 
 
\section{Conclusions}

In this work, we have presented a novel extension of RT-TDDFTB dynamics to including e-e scattering effects in metal nanoparticles.
Our new approach combines the Fermi's golden rule style relaxation rate in the quantum Boltzmann equation (QBE) with a Lindbladian form of the dissipation operator. The interaction coefficients are evaluated from the RPA screened potential based on the DFTB Coulomb matrix, which enables a self-consistent description of the electron dynamics from the initial plasmonic excitation to the electron relaxation without any phenomenological parameters. Furthermore, as a general extension, we expect our theory to be compatible with regular TDDFT as well.

By applying our RPA-LQBE approach to the RT-TDDFTB modeling of silver, gold and aluminum icosahedral nanoclusters with 147-561 atoms, we have investigated the quasiparticle lifetime, as well as the population and coherence relaxation after a laser pulse. 
Our calculated quasiparticle lifetime generally agrees with the existing results. Despite the small size of the clusters, the quasiparticle lifetimes are close to the bulk value, although with large fluctuations near the Fermi surface.

In our dynamics simulations, the energetic hot carriers (> 1 eV) relax efficiently in almost all clusters within 100 fs after the laser pulse, and the population distributions are already close to a high temperature Boltzmann distribution at 300 fs. 
The relaxation dynamics show very strong energy-dependence. For the total population, the relaxation times in gold and silver are around 100-400 fs and 100-200 fs, respectively, consistent with typical experimental values. For the energetic hot electrons (> 1 eV), the relaxation time is below 150 fs for gold and below 100 fs for silver. This observation indicates the importance of an energy-specific treatment when modeling electron relaxation.
For small clusters, quantum effects can also lead to the deviations of relaxation dynamics from the typical thermalization picture. For gold, due to the high energy of the 5d-band, the Auger electrons generated from hole scattering can slow down the relaxation of hot electrons > 1 eV.

We have observed fast decoherence associated with plasmonic resonance excitation in small clusters at a timescale of 10 fs, significantly faster than the population equilibration. For gold clusters, we also observe a secondary decoherence process from interband excitations at a longer timescale.

Overall, our findings indicate that e-e scattering is an important ingredient for accurate modeling of plasmonic dynamics, and the gold 5d-band has a strong influence in the hot carrier relaxation.
Since DFTB is orders of magnitude faster than conventional TDDFT, this suggests that DFTB-based theory provides an important capability for characterizing electron dynamics for nanoclusters that spans the gap from molecular to nanoparticle behavior. 

\section{Methods}

\subsection{Real-Time TDDFTB}

In the MO basis, the RT-TDDFTB equation of motion is identical to the RT-TDDFT equation (Eq.~\eqref{eq:rttddft}) except $h^{\text{DFT}}$ is replaced by the DFTB Hamiltonian $h^{\text{DFTB}}$, which is given by
\begin{align}
    h_{pp'}^{\text{DFTB}} = h^{0,\text{DFTB}}_{pp'} + \sum_{q,q'}{v^{\text{DFTB}}_{pp',qq'}\rho_{qq'}} \label{eq:dftb}
\end{align}
In the SCC-DFTB formalism, the terms in Eq.~\eqref{eq:dftb} are given by \cite{elstner1998Selfconsistentcharge} 
\begin{align}
    h^{0,\text{DFTB}}_{pp'} &= \sum_{\mu,\nu}{C_{\mu p}C_{\nu p'}t_{\mu\nu}} - \sum_{\mu,\nu,I,J}{S_{\mu\nu}^Jq_0^I\gamma_{IJ}C_{\mu p}C_{\nu p'}} \\
    v^{\text{DFTB}}_{pp',qq'} &= \sum_{\mu,\nu,\sigma,\tau,I,J}{C_{\mu p}C_{\nu p'}C_{\sigma q}C_{\tau q'}S_{\mu\nu}^IS_{\sigma\tau}^J\gamma_{IJ}} \\
    S_{\mu\nu}^I&=\frac 12(S_{\mu\nu}|_{\mu\in\text{atom }I} + S_{\mu\nu}|_{\nu\in\text{atom }I})
\end{align}
where $t_{\mu\nu}$ is the first-order tight-binding element between atomic orbitals $\mu$ and $\nu$, with $\gamma_{IJ}$ is the Coulomb + XC matrix element between atoms $I,J$, $C_{\mu p}$ is the MO coefficient and $S_{\mu\nu}$ is the overlap matrix. The SCC-DFTB Hamiltonian can be viewed as a dramatic simplification to the DFT one, with a factorized Coulomb and time-independent XC kernel.

\subsection{RPA Screened Coulomb with DFTB}

To evaluate the screened Coulomb matrix $W$ appeared in Eq.~\eqref{eq:Gamma}, we use the random-phase approximation (RPA) equation with the DFTB Coulomb+XC operator:\cite{echenique2000Theory}
\begin{align}
    W_{pq,rs}(\omega) = v^{\text{DFTB}}_{pq,rs} + \sum_{m,n,m',n'}{v^{\text{DFTB}}_{pq,mn}\Pi_{mn,m'n'}(\omega)W_{m'n',rs}(\omega)} \label{eq:w}
\end{align}
where 
\begin{align}
    \Pi_{mn,m'n'}(\omega) = 2\frac{f_m-f_n}{\hbar(\omega - \omega_m + \omega_n + i\gamma_0)}\delta_{mm'}\delta_{nn'} \label{eq:Pi}
\end{align}
is the Kohn-Sham (KS) response function.

Since $v^{\text{DFTB}}$ contains the XC interaction, the result differs from the exact RPA value where $v$ is the pure Coulomb potential. However, for metals the XC part is generally small and localized, therefore we expect the difference to be no more than 10\%.

In principle, the screened Coulomb potential can be solved by a matrix inversion:
\begin{align}
    \mathbf{W}(\omega) = (\mathbf{1} - \mathbf{v}\bm{\Pi}(\omega))^{-1}\mathbf{v} \label{eq:w_inv}
\end{align}
where the size of each matrix is $N^2\times N^2$. For a large system, Eq.~\eqref{eq:w_inv} is hardly solvable due to the extreme computational cost. Nevertheless, with DFTB, the interaction potential can be efficiently factorized:
\begin{align}
    v^{\text{DFTB}}_{pp',qq'} &= \sum_{\mu,\nu,\sigma,\tau,I,J}{C_{\mu p}C_{\nu p'}C_{\sigma q}C_{\tau q'}S_{\mu\nu}^IS_{\sigma\tau}^J\gamma_{IJ}}  \equiv \sum_{I,J}{Q_{pp'}^IQ_{qq'}^J\gamma_{IJ}}
\end{align}
where $Q^I_{pp'}=\sum_{\mu,\nu}{C_{\mu p}C_{\nu p'}S_{\mu\nu}^I}$ is the orbital resolved Mulliken charge. By a repeated expansion of Eq.~\eqref{eq:w}, we now have
\begin{align}
    W_{pq,rs} &= v^{\text{DFTB}}_{pq,rs} + \sum_{m,n,m',n'}{v^{\text{DFTB}}_{pq,mn}\Pi_{mn,m'n'}v^{\text{DFTB}}_{m'n',rs}} + ... \nonumber\\
     &= \sum_{I,J}{Q_{pq}^IQ_{rs}^J\gamma_{IJ}} + \sum_{m,n,m',n',I,J,K,L}{Q_{pq}^IQ_{mn}^K\gamma_{IK}\Pi_{mn,m'n'}Q_{m'n'}^{L}Q_{rs}^J\gamma_{LJ}} + ... \nonumber\\
     &= \sum_{I,J}{Q_{pq}^IQ_{rs}^J\left(\gamma_{IJ} + \sum_{m,n,m',n',K,L}{\gamma_{IK}(Q_{mn}^K\Pi_{mn,m'n'}Q_{m'n'}^{L})\gamma_{LJ}} + ...\right)}
\end{align}
By defining the atom-resolved KS response matrix
\begin{align}
    \pi_{IJ}(\omega) = \sum_{m,n,m',n'}{Q^I_{mn}Q^J_{m'n'}\Pi_{mn,m'n'}(\omega)} \label{eq:pi}
\end{align}
, we finally arrive at
\begin{align}
    W_{pq,rs}(\omega) &= \sum_{I,J}{Q_{pq}^IQ_{rs}^J(\gamma_{IJ} + \sum_{K,L}{\gamma_{IK}\pi_{KL}(\omega)\gamma_{LJ}} + ...)} \nonumber\\
    &= \sum_{I,J}{Q_{pq}^IQ_{rs}^J(\sum_{K}{(1-\gamma\pi(\omega))^{-1}_{IK}\gamma_{KJ}})} \label{eq:w_atom}
\end{align}
In Eq.~\eqref{eq:w_atom}, the matrix size to be inverted is only $N_{\text{atom}}\times N_{\text{atom}}$, which greatly improves the efficiency. In our implementation, the $\pi$ matrix in Eq.~\eqref{eq:pi} is pre-computed and stored before the dynamics simulations.

\subsection{Computational Details}

The geometries are built using the Atomic Simulation Environment (ASE) package \cite{larsen2017atomic} with lattice constant $l_{\text{Ag}}=$ 4.0853 Angstrom, $l_{\text{Au}}=$ 4.0782 Angstrom, and $l_{\text{Al}}=$ 4.0495 Angstrom (see SI for the XYZ files). The nuclear geometries are not optimized and are frozen throughout the simulation.

The ground state calculations were calculated with DFTB+ package.\cite{hourahine2020DFTB} For the DFTB Slater-Koster parameters, we use Hyb-0-2 for silver, Auorg-1-1 \cite{fihey2015SCCDFTB} for gold and Matsci-0-3 \cite{frenzel2004Semirelativistic} for aluminum. 

For the real-time calculation, the density matrix $\rho$ is propagated by a leapfrog integrator with a timestep of 0.1 au (0.0024 fs) except for \ch{Al_{147}} we use 0.05 au (0.0012 fs). The $\bar{\rho}$ in Eq.~(7) is updated each step after $\rho$ is obtained. Specifically, for $p\ne q$, we use the standard implementation of the low pass filter:
\begin{align}
    \bar{\rho}_{pq}(t+\Delta t) = e^{-i\omega_{pq}\Delta t-\gamma\Delta t}\bar{\rho}_{pq}(t) + (1-e^{-\gamma\Delta t})\rho_{pq}(t+\Delta t)
\end{align}
with $\hbar\gamma=\text{1 eV}$ (see Sec.~S2A in the SI).

For the LQBE terms, the screened potential is evaluated once with the ground state electronic configuration before the simulation starts, on a frequency grid between 0-6 eV with a spacing of 0.01 eV. During the dynamics simulations, the potential is linearly interpolated at $\omega_{pq}$ when evaluating $\Gamma_{pqrs}$ (see Eq.~\eqref{eq:Gamma}). The $\gamma_0$ in Eq.~\eqref{eq:Pi} is chosen to be $\text{100 fs}^{-1}$ for all systems and has little effect on the result. 

To reduce the computational cost, we utilize the commonly used linearization scheme of QBE, where only one of the population components (the ``active'' component) are updated each step. To address the long time dynamics, instead of using the equilibrium value, the other three ``inactive'' components are also updated every 200 au (4.8 fs). The update frequency has very little effect on the dynamics. Details of linearization are described in Sec.~S1C in the SI.

Due to the finite size and high symmetry of our systems, the orbital energies are often degenerate and a large amount of scattering can occur between these degenerate orbitals (i.e. $\omega_p=\omega_q$, $\omega_p=\omega_s$, $\omega_r=\omega_q$ or $\omega_r=\omega_s$ in Eq.~\eqref{eq:Gamma}), especially for orbitals around the Fermi surface where the number of scattering channels is limited. These elastic and nearly elastic scattering processes are strongly coherent and do not follow the Fermi golden rule (FGR) dissipation rate (they are nonperturbative and more like Rabi oscillations instead). Due to the small energy gap associated with elastic scattering, the standard QBE (which is Markovian) would treat them as incoherent and overestimate the elastic scattering rate. As a simplest treatment, we exclude these elastic processes in our calculations, which turns out to have little effect on the inelastic dynamics of interest in this work, but substantially improves the lifetime profile (for a detailed analysis, see Sec.~S2B in the SI).

For the real-time dynamics, we choose $\hbar\omega_{\text{exc}}$ to be 2.89 eV, 2.84 eV and 2.76 eV for \ch{Ag_{147}}, \ch{Ag_{309}} and \ch{Ag_{561}}, 2.24 eV, 2.33 eV and 2.30 eV for \ch{Au_{147}}, \ch{Au_{309}} and \ch{Au_{561}}, and 5.0 eV for all aluminum clusters.

Due to the very weak response, for the $\hbar\omega_{\text{exc}}=\text{1.5 eV}$ simulations, we use a strong $E_0=\text{0.05 V/Angstrom}$ instead 0.01 V/Angstrom. The population dynamics are unaffected as the strength is still in the linear response region.

For spectral calculations, the system is perturbed by a delta-kick with $E_0=$0.001 V/Angstrom. For the spectra in Fig.~\ref{fig:spec}, we add a 20 fs Lorentzian broadening for post-processing. 
For the spectra in Fig.~\ref{fig:deco}a, we add a 100 fs Lorentzian broadening to the spectra for better visualization (note that this is much narrower than the typical spectral simulation). The artificial broadening is applied to both standard RT-TDDFTB and RT-TDDFTB+LQBE in Fig.~\ref{fig:deco}a to keep the total width comparable. Note that for RT-TDDFTB+LQBE, the spectra are already smooth without any artificial broadening.

\subsection{Relaxation Time Fitting}

To skip the irregular pattern during the laser pulse and rising stage of the population between 1-2 eV, we use the population data between 30 and 300 fs relative to the laser peak time for the fitting shown in Fig.~\ref{fig:pop}. The fit results for all clusters are shown in Fig.~S18 in the SI.

\section*{Supporting Information}

Supporting Information: Figures S1-S18, Notes S1-S3 deriving equations and supporting reference (PDF). 

All geometries of the clusters in XYZ format (ZIP).

\section*{Acknowledgment}
This research was supported by the Division of Chemistry of the National Science Foundation under grant CHE-2347622. 

\bibliography{Plasmon}

@incollection{1991Landau,
  title = {Landau Fermi-Liquid Theory and Low Temperature Properties of Normal Liquid 3he},
  booktitle = {Landau Fermi-liquid Theory},
  year = 1991,
  eprint = {https://onlinelibrary.wiley.com/doi/pdf/10.1002/9783527617159.ch1},
  pages = {1--121},
  publisher = {John Wiley \& Sons, Ltd},
  doi = {10.1002/9783527617159.ch1},
  chapter = {1},
  isbn = {978-3-527-61715-9}
}

@article{ahlawat2021Plasmoninduced,
  title = {Plasmon-Induced Hot-Hole Generation and Extraction at Nano-Heterointerfaces for Photocatalysis},
  author = {Ahlawat, Monika and Mittal, Diksha and Govind Rao, Vishal},
  year = 2021,
  month = nov,
  journal = {Communications Materials},
  volume = {2},
  number = {1},
  pages = {114},
  issn = {2662-4443},
  doi = {10.1038/s43246-021-00220-4},
  urldate = {2026-01-22},
  langid = {english}
}

@article{aikens2008Discrete,
  title = {From {{Discrete Electronic States}} to {{Plasmons}}: {{TDDFT Optical Absorption Properties}} of {{Ag}}{\textsubscript{ {\emph{n}} }} ( {\emph{n}} = 10, 20, 35, 56, 84, 120) {{Tetrahedral Clusters}}},
  shorttitle = {From {{Discrete Electronic States}} to {{Plasmons}}},
  author = {Aikens, Christine M. and Li, Shuzhou and Schatz, George C.},
  year = 2008,
  month = jul,
  journal = {The Journal of Physical Chemistry C},
  volume = {112},
  number = {30},
  pages = {11272--11279},
  issn = {1932-7447, 1932-7455},
  doi = {10.1021/jp802707r},
  urldate = {2026-02-06},
  langid = {english}
}

@article{asadi-aghbolaghi2020TDDFT+TB,
  title = {{{TD-DFT}}+{{TB}}: {{An Efficient}} and {{Fast Approach}} for {{Quantum Plasmonic Excitations}}},
  shorttitle = {{{TD-DFT}}+{{TB}}},
  author = {{Asadi-Aghbolaghi}, Narges and R{\"u}ger, Robert and Jamshidi, Zahra and Visscher, Lucas},
  year = 2020,
  month = apr,
  journal = {The Journal of Physical Chemistry C},
  volume = {124},
  number = {14},
  pages = {7946--7955},
  issn = {1932-7447, 1932-7455},
  doi = {10.1021/acs.jpcc.0c00979},
  urldate = {2025-12-04},
  copyright = {http://pubs.acs.org/page/policy/authorchoice\_ccbyncnd\_termsofuse.html},
  langid = {english}
}

@inproceedings{bauer1998Electron,
  title = {Electron Dynamics of Aluminum Investigated by Means of Time-Resolved Photoemission},
  booktitle = {Laser Techniques for Surface Science {{III}}},
  author = {Bauer, M and Pawlik, S and Aeschlimann, Martin},
  year = 1998,
  volume = {3272},
  pages = {201--210},
  publisher = {SPIE}
}

@article{bauer2015Hot,
  title = {Hot Electron Lifetimes in Metals Probed by Time-Resolved Two-Photon Photoemission},
  author = {Bauer, M. and Marienfeld, A. and Aeschlimann, M.},
  year = 2015,
  month = aug,
  journal = {Progress in Surface Science},
  volume = {90},
  number = {3},
  pages = {319--376},
  issn = {00796816},
  doi = {10.1016/j.progsurf.2015.05.001},
  urldate = {2025-12-04},
  langid = {english}
}

@article{baumgratz2014Quantifying,
  title = {Quantifying Coherence},
  author = {Baumgratz, T. and Cramer, M. and Plenio, M. B.},
  year = 2014,
  month = sep,
  journal = {Physical Review Letters},
  volume = {113},
  number = {14},
  pages = {140401},
  publisher = {American Physical Society},
  doi = {10.1103/PhysRevLett.113.140401}
}

@article{berdakin2020Interplay,
  title = {Interplay between {{Intra-}} and {{Interband Transitions Associated}} with the {{Plasmon-Induced Hot Carrier Generation Process}} in {{Silver}} and {{Gold Nanoclusters}}},
  author = {Berdakin, Matias and {Douglas-Gallardo}, Oscar A. and S{\'a}nchez, Cristi{\'a}n G.},
  year = 2020,
  month = jan,
  journal = {The Journal of Physical Chemistry C},
  volume = {124},
  number = {2},
  pages = {1631--1639},
  issn = {1932-7447, 1932-7455},
  doi = {10.1021/acs.jpcc.9b10871},
  urldate = {2026-01-23},
  copyright = {https://doi.org/10.15223/policy-029},
  langid = {english}
}

@article{bernardi2015Theory,
  title = {Theory and Computation of Hot Carriers Generated by Surface Plasmon Polaritons in Noble Metals},
  author = {Bernardi, Marco and Mustafa, Jamal and Neaton, Jeffrey B. and Louie, Steven G.},
  year = 2015,
  month = jun,
  journal = {Nature Communications},
  volume = {6},
  number = {1},
  pages = {7044},
  issn = {2041-1723},
  doi = {10.1038/ncomms8044},
  urldate = {2025-12-04},
  langid = {english}
}

@article{besteiro2017Understanding,
  title = {Understanding {{Hot-Electron Generation}} and {{Plasmon Relaxation}} in {{Metal Nanocrystals}}: {{Quantum}} and {{Classical Mechanisms}}},
  shorttitle = {Understanding {{Hot-Electron Generation}} and {{Plasmon Relaxation}} in {{Metal Nanocrystals}}},
  author = {Besteiro, Lucas V. and Kong, Xiang-Tian and Wang, Zhiming and Hartland, Gregory and Govorov, Alexander O.},
  year = 2017,
  month = nov,
  journal = {ACS Photonics},
  volume = {4},
  number = {11},
  pages = {2759--2781},
  issn = {2330-4022, 2330-4022},
  doi = {10.1021/acsphotonics.7b00751},
  urldate = {2025-12-04},
  langid = {english}
}

@article{bhasin2025Plasmon,
  title = {Plasmon {{Dynamics}} in {{Nanoclusters}}: {{Dephasing Revealed}} by {{Excited States Evaluation}}},
  shorttitle = {Plasmon {{Dynamics}} in {{Nanoclusters}}},
  author = {Bhasin, Anant O. and Ceylan, Yavuz S. and Dillon, Alva D. and Giri, Sajal Kumar and Schatz, George C. and Gieseking, Rebecca L. M.},
  year = 2025,
  month = jan,
  journal = {Journal of Chemical Theory and Computation},
  volume = {21},
  number = {1},
  pages = {17--28},
  issn = {1549-9618, 1549-9626},
  doi = {10.1021/acs.jctc.4c01302},
  urldate = {2025-12-04},
  copyright = {https://doi.org/10.15223/policy-029},
  langid = {english}
}

@article{bigot2000Electron,
  title = {Electron Dynamics in Metallic Nanoparticles},
  author = {Bigot, J.-Y. and Halt{\'e}, V. and Merle, J.-C. and Daunois, A.},
  year = 2000,
  month = jan,
  journal = {Chemical Physics},
  volume = {251},
  number = {1-3},
  pages = {181--203},
  issn = {03010104},
  doi = {10.1016/S0301-0104(99)00298-0},
  urldate = {2025-12-13},
  copyright = {https://www.elsevier.com/tdm/userlicense/1.0/},
  langid = {english}
}

@article{binder1992Carriercarrier,
  title = {Carrier-Carrier Scattering and Optical Dephasing in Highly Excited Semiconductors},
  author = {Binder, R. and Scott, D. and Paul, A. E. and Lindberg, M. and Henneberger, K. and Koch, S. W.},
  year = 1992,
  month = jan,
  journal = {Physical Review B},
  volume = {45},
  number = {3},
  pages = {1107--1115},
  issn = {0163-1829, 1095-3795},
  doi = {10.1103/PhysRevB.45.1107},
  urldate = {2025-12-06},
  copyright = {http://link.aps.org/licenses/aps-default-license},
  langid = {english}
}

@article{bohne1990Phase,
  title = {Phase Memory of the Electronic Polarization in Transient Nonlinear Optical Spectra of Gallium Arsenide at 2 {{eV}}},
  author = {B{\"o}hne, G. and Sure, T. and Ulbrich, R. G. and Sch{\"a}fer, W.},
  year = 1990,
  month = apr,
  journal = {Physical Review B},
  volume = {41},
  number = {11},
  pages = {7549--7558},
  issn = {0163-1829, 1095-3795},
  doi = {10.1103/PhysRevB.41.7549},
  urldate = {2025-12-06},
  copyright = {http://link.aps.org/licenses/aps-default-license},
  langid = {english}
}

@article{bonitz1996Numerical,
  title = {Numerical Analysis of Non-{{Markovian}} Effects in Charge-Carrier Scattering: One-Time versus Two-Time Kinetic Equations},
  shorttitle = {Numerical Analysis of Non-{{Markovian}} Effects in Charge-Carrier Scattering},
  author = {Bonitz, M and Kremp, D and Scott, D C and Binder, R and Kraeft, W D and K{\"o}hler, H S},
  year = 1996,
  month = aug,
  journal = {Journal of Physics: Condensed Matter},
  volume = {8},
  number = {33},
  pages = {6057--6071},
  issn = {0953-8984, 1361-648X},
  doi = {10.1088/0953-8984/8/33/012},
  urldate = {2025-12-04},
  langid = {english}
}

@book{bonitz2016Quantum,
  title = {Quantum {{Kinetic Theory}}},
  author = {Bonitz, Michael},
  year = 2016,
  publisher = {Springer International Publishing},
  address = {Cham},
  doi = {10.1007/978-3-319-24121-0},
  urldate = {2025-12-06},
  copyright = {http://www.springer.com/tdm},
  isbn = {978-3-319-24119-7 978-3-319-24121-0},
  langid = {english}
}

@article{brown2016Nonradiative,
  title = {Nonradiative {{Plasmon Decay}} and {{Hot Carrier Dynamics}}: {{Effects}} of {{Phonons}}, {{Surfaces}}, and {{Geometry}}},
  shorttitle = {Nonradiative {{Plasmon Decay}} and {{Hot Carrier Dynamics}}},
  author = {Brown, Ana M. and Sundararaman, Ravishankar and Narang, Prineha and Goddard, William A. and Atwater, Harry A.},
  year = 2016,
  month = jan,
  journal = {ACS Nano},
  volume = {10},
  number = {1},
  pages = {957--966},
  issn = {1936-0851, 1936-086X},
  doi = {10.1021/acsnano.5b06199},
  urldate = {2025-12-04},
  langid = {english}
}

@article{brown2017Experimental,
  title = {Experimental and {{{\emph{Ab Initio}}}} {{Ultrafast Carrier Dynamics}} in {{Plasmonic Nanoparticles}}},
  author = {Brown, Ana M. and Sundararaman, Ravishankar and Narang, Prineha and Schwartzberg, Adam M. and Goddard, William A. and Atwater, Harry A.},
  year = 2017,
  month = feb,
  journal = {Physical Review Letters},
  volume = {118},
  number = {8},
  pages = {087401},
  issn = {0031-9007, 1079-7114},
  doi = {10.1103/PhysRevLett.118.087401},
  urldate = {2025-12-04},
  copyright = {http://link.aps.org/licenses/aps-default-license},
  langid = {english}
}

@article{campillo2000Hole,
  title = {Hole {{Dynamics}} in {{Noble Metals}}},
  author = {Campillo, I. and Rubio, A. and Pitarke, J. M. and Goldmann, A. and Echenique, P. M.},
  year = 2000,
  month = oct,
  journal = {Physical Review Letters},
  volume = {85},
  number = {15},
  pages = {3241--3244},
  issn = {0031-9007, 1079-7114},
  doi = {10.1103/PhysRevLett.85.3241},
  urldate = {2025-12-06},
  copyright = {http://link.aps.org/licenses/aps-default-license},
  langid = {english}
}

@article{casida2012Progress,
  title = {Progress in {{Time-Dependent Density-Functional Theory}}},
  author = {Casida, M.E. and {Huix-Rotllant}, M.},
  year = 2012,
  month = may,
  journal = {Annual Review of Physical Chemistry},
  volume = {63},
  number = {1},
  pages = {287--323},
  issn = {0066-426X, 1545-1593},
  doi = {10.1146/annurev-physchem-032511-143803},
  urldate = {2025-12-17},
  langid = {english}
}

@article{cazzaniga2012approaches,
  title = {G {{W}} and beyond Approaches to Quasiparticle Properties in Metals},
  author = {Cazzaniga, Marco},
  year = 2012,
  month = jul,
  journal = {Physical Review B},
  volume = {86},
  number = {3},
  pages = {035120},
  issn = {1098-0121, 1550-235X},
  doi = {10.1103/PhysRevB.86.035120},
  urldate = {2025-12-22},
  copyright = {http://link.aps.org/licenses/aps-default-license},
  langid = {english}
}

@article{chellam2025Density,
  title = {Density {{Functional Tight-Binding Captures Plasmon-Driven H}}{\textsubscript{2}} {{Dissociation}} on {{Al Nanocrystals}}},
  author = {Chellam, Nikhil S. and Schatz, George C.},
  year = 2025,
  month = may,
  journal = {The Journal of Physical Chemistry C},
  volume = {129},
  number = {18},
  pages = {8634--8644},
  issn = {1932-7447, 1932-7455},
  doi = {10.1021/acs.jpcc.5c00805},
  urldate = {2025-12-04},
  copyright = {https://doi.org/10.15223/policy-029},
  langid = {english}
}

@article{cuny2018Densityfunctional,
  title = {Density-Functional Tight-Binding Approach for Metal Clusters, Nanoparticles, Surfaces and Bulk: Application to Silver and Gold},
  shorttitle = {Density-Functional Tight-Binding Approach for Metal Clusters, Nanoparticles, Surfaces and Bulk},
  author = {Cuny, J{\'e}r{\^o}me and Tarrat, Nathalie and Spiegelman, Fernand and Huguenot, Arthur and Rapacioli, Mathias},
  year = 2018,
  month = aug,
  journal = {Journal of Physics: Condensed Matter},
  volume = {30},
  number = {30},
  pages = {303001},
  issn = {0953-8984, 1361-648X},
  doi = {10.1088/1361-648X/aacd6c},
  urldate = {2025-12-04},
  langid = {english}
}

@article{dagostino2018Density,
  title = {Density {{Functional Tight Binding}} for {{Quantum Plasmonics}}},
  author = {D'Agostino, Stefania and Rinaldi, Rosaria and Cuniberti, Gianaurelio and Della Sala, Fabio},
  year = 2018,
  month = aug,
  journal = {The Journal of Physical Chemistry C},
  volume = {122},
  number = {34},
  pages = {19756--19766},
  issn = {1932-7447, 1932-7455},
  doi = {10.1021/acs.jpcc.8b05278},
  urldate = {2025-12-04},
  langid = {english}
}

@article{dallosto2024Peeking,
  title = {Peeking into the {{Femtosecond Hot-Carrier Dynamics Reveals Unexpected Mechanisms}} in {{Plasmonic Photocatalysis}}},
  author = {Dall'Osto, Giulia and Marsili, Margherita and Vanzan, Mirko and Toffoli, Daniele and Stener, Mauro and Corni, Stefano and Coccia, Emanuele},
  year = 2024,
  month = jan,
  journal = {Journal of the American Chemical Society},
  volume = {146},
  number = {3},
  pages = {2208--2218},
  issn = {0002-7863, 1520-5126},
  doi = {10.1021/jacs.3c12470},
  urldate = {2025-12-04},
  copyright = {https://creativecommons.org/licenses/by/4.0/},
  langid = {english}
}

@article{douglas-gallardo2019Plasmoninduced,
  title = {Plasmon-Induced Hot-Carrier Generation Differences in Gold and Silver Nanoclusters},
  author = {{Douglas-Gallardo}, Oscar A. and Berdakin, Mat{\'i}as and Frauenheim, Thomas and S{\'a}nchez, Cristi{\'a}n G.},
  year = 2019,
  journal = {Nanoscale},
  volume = {11},
  number = {17},
  pages = {8604--8615},
  issn = {2040-3364, 2040-3372},
  doi = {10.1039/C9NR01352K},
  urldate = {2025-12-04},
  langid = {english}
}

@article{echenique2000Theory,
  title = {Theory of Inelastic Lifetimes of Low-Energy Electrons in Metals},
  author = {Echenique, P.M. and Pitarke, J.M. and Chulkov, E.V. and Rubio, A.},
  year = 2000,
  month = jan,
  journal = {Chemical Physics},
  volume = {251},
  number = {1-3},
  pages = {1--35},
  issn = {03010104},
  doi = {10.1016/S0301-0104(99)00313-4},
  urldate = {2025-12-04},
  copyright = {https://www.elsevier.com/tdm/userlicense/1.0/},
  langid = {english}
}

@article{elstner1998Selfconsistentcharge,
  title = {Self-Consistent-Charge Density-Functional Tight-Binding Method for Simulations of Complex Materials Properties},
  author = {Elstner, M. and Porezag, D. and Jungnickel, G. and Elsner, J. and Haugk, M. and Frauenheim, {\relax Th}. and Suhai, S. and Seifert, G.},
  year = 1998,
  month = sep,
  journal = {Physical Review B},
  volume = {58},
  number = {11},
  pages = {7260--7268},
  issn = {0163-1829, 1095-3795},
  doi = {10.1103/PhysRevB.58.7260},
  urldate = {2025-12-04},
  copyright = {http://link.aps.org/licenses/aps-default-license},
  langid = {english}
}

@article{faleev2004AllElectron,
  title = {All-{{Electron Self-Consistent G W Approximation}}: {{Application}} to {{Si}}, {{MnO}}, and {{NiO}}},
  shorttitle = {All-{{Electron Self-Consistent G W Approximation}}},
  author = {Faleev, Sergey V. and Van Schilfgaarde, Mark and Kotani, Takao},
  year = 2004,
  month = sep,
  journal = {Physical Review Letters},
  volume = {93},
  number = {12},
  pages = {126406},
  issn = {0031-9007, 1079-7114},
  doi = {10.1103/PhysRevLett.93.126406},
  urldate = {2025-12-22},
  copyright = {http://link.aps.org/licenses/aps-default-license},
  langid = {english}
}

@article{fihey2015SCCDFTB,
  title = {{{SCC-DFTB}} Parameters for Simulating Hybrid Gold-Thiolates Compounds},
  author = {Fihey, Arnaud and Hettich, Christian and Touzeau, J{\'e}r{\'e}my and Maurel, Fran{\c c}ois and Perrier, Aur{\'e}lie and K{\"o}hler, Christof and Aradi, B{\'a}lint and Frauenheim, Thomas},
  year = 2015,
  journal = {Journal of Computational Chemistry},
  volume = {36},
  number = {27},
  pages = {2075--2087},
  publisher = {Wiley Online Library}
}

@article{frenzel2004Semirelativistic,
  title = {Semi-Relativistic, Self-Consistent Charge {{Slater-Koster}} Tables for Density-Functional Based Tight-Binding ({{DFTB}}) for Materials Science Simulations},
  author = {Frenzel, Johannes and Oliveira, {\relax AF} and Jardillier, N and Heine, T and Seifert, G},
  year = 2004,
  journal = {Zeolites},
  volume = {2},
  number = {3},
  pages = {7}
}

@article{giannone2020Minimal,
  title = {Minimal Auxiliary Basis Set for Time-Dependent Density Functional Theory and Comparison with Tight-Binding Approximations: {{Application}} to Silver Nanoparticles},
  shorttitle = {Minimal Auxiliary Basis Set for Time-Dependent Density Functional Theory and Comparison with Tight-Binding Approximations},
  author = {Giannone, Giulia and Della Sala, Fabio},
  year = 2020,
  month = aug,
  journal = {The Journal of Chemical Physics},
  volume = {153},
  number = {8},
  pages = {084110},
  issn = {0021-9606, 1089-7690},
  doi = {10.1063/5.0020545},
  urldate = {2025-12-04},
  langid = {english}
}

@article{giri2023Photodissociation,
  title = {Photodissociation of {{H}}{\textsubscript{2}} on {{Ag}} and {{Au Nanoparticles}}: {{Effect}} of {{Size}} and {{Plasmon}} versus {{Interband Transitions}} on {{Threshold Intensities}} for {{Dissociation}}},
  shorttitle = {Photodissociation of {{H}}{\textsubscript{2}} on {{Ag}} and {{Au Nanoparticles}}},
  author = {Giri, Sajal Kumar and Schatz, George C.},
  year = 2023,
  month = mar,
  journal = {The Journal of Physical Chemistry C},
  volume = {127},
  number = {8},
  pages = {4115--4123},
  issn = {1932-7447, 1932-7455},
  doi = {10.1021/acs.jpcc.3c00006},
  urldate = {2025-12-04},
  copyright = {https://doi.org/10.15223/policy-029},
  langid = {english}
}

@article{golze2019GW,
  title = {The {{GW Compendium}}: {{A Practical Guide}} to {{Theoretical Photoemission Spectroscopy}}},
  shorttitle = {The {{GW Compendium}}},
  author = {Golze, Dorothea and Dvorak, Marc and Rinke, Patrick},
  year = 2019,
  month = jul,
  journal = {Frontiers in Chemistry},
  volume = {7},
  pages = {377},
  issn = {2296-2646},
  doi = {10.3389/fchem.2019.00377},
  urldate = {2025-12-04},
  langid = {english}
}

@article{govorov2015Kinetic,
  title = {Kinetic {{Density Functional Theory}} for {{Plasmonic Nanostructures}}: {{Breaking}} of the {{Plasmon Peak}} in the {{Quantum Regime}} and {{Generation}} of {{Hot Electrons}}},
  shorttitle = {Kinetic {{Density Functional Theory}} for {{Plasmonic Nanostructures}}},
  author = {Govorov, Alexander O. and Zhang, Hui},
  year = 2015,
  month = mar,
  journal = {The Journal of Physical Chemistry C},
  volume = {119},
  number = {11},
  pages = {6181--6194},
  issn = {1932-7447, 1932-7455},
  doi = {10.1021/jp512105m},
  urldate = {2025-12-04},
  langid = {english}
}

@article{grimme2016Ultrafast,
  title = {Ultra-Fast Computation of Electronic Spectra for Large Systems by Tight-Binding Based Simplified {{Tamm-Dancoff}} Approximation ({{sTDA-xTB}})},
  author = {Grimme, Stefan and Bannwarth, Christoph},
  year = 2016,
  month = aug,
  journal = {The Journal of Chemical Physics},
  volume = {145},
  number = {5},
  pages = {054103},
  issn = {0021-9606, 1089-7690},
  doi = {10.1063/1.4959605},
  urldate = {2025-12-04},
  langid = {english}
}

@article{gross1985Local,
  title = {Local Density-Functional Theory of Frequency-Dependent Linear Response},
  author = {Gross, {\relax EKU} and Kohn, Walter},
  year = 1985,
  journal = {Physical Review Letters},
  volume = {55},
  number = {26},
  pages = {2850},
  publisher = {APS}
}

@article{haas1996Generalized,
  title = {Generalized {{Monte Carlo}} Approach for the Study of the Coherent Ultrafast Carrier Dynamics in Photoexcited Semiconductors},
  author = {Haas, Stefan and Rossi, Fausto and Kuhn, Tilmann},
  year = 1996,
  month = may,
  journal = {Physical Review B},
  volume = {53},
  number = {19},
  pages = {12855--12868},
  issn = {0163-1829, 1095-3795},
  doi = {10.1103/PhysRevB.53.12855},
  urldate = {2025-12-06},
  copyright = {http://link.aps.org/licenses/aps-default-license},
  langid = {english}
}

@article{habib2023Machine,
  title = {Machine {{Learning Models Capture Plasmon Dynamics}} in {{Ag Nanoparticles}}},
  author = {Habib, Adela and Lubbers, Nicholas and Tretiak, Sergei and Nebgen, Benjamin},
  year = 2023,
  month = may,
  journal = {The Journal of Physical Chemistry A},
  volume = {127},
  number = {17},
  pages = {3768--3778},
  issn = {1089-5639, 1520-5215},
  doi = {10.1021/acs.jpca.2c08757},
  urldate = {2025-12-16},
  copyright = {https://creativecommons.org/licenses/by-nc-nd/4.0/},
  langid = {english}
}

@article{herring2023Mechanistic,
  title = {Mechanistic {{Insights}} into {{Plasmonic Catalysis}} by {{Dynamic Calculations}}: {{O}}{\textsubscript{2}} and {{N}}{\textsubscript{2}} on {{Au}} and {{Ag Nanoparticles}}},
  shorttitle = {Mechanistic {{Insights}} into {{Plasmonic Catalysis}} by {{Dynamic Calculations}}},
  author = {Herring, Connor J. and Montemore, Matthew M.},
  year = 2023,
  month = feb,
  journal = {Chemistry of Materials},
  volume = {35},
  number = {4},
  pages = {1586--1593},
  issn = {0897-4756, 1520-5002},
  doi = {10.1021/acs.chemmater.2c03061},
  urldate = {2025-12-04},
  copyright = {https://creativecommons.org/licenses/by/4.0/},
  langid = {english}
}

@article{herring2023Recent,
  title = {Recent {{Advances}} in {{Real-Time Time-Dependent Density Functional Theory Simulations}} of {{Plasmonic Nanostructures}} and {{Plasmonic Photocatalysis}}},
  author = {Herring, Connor J. and Montemore, Matthew M.},
  year = 2023,
  month = aug,
  journal = {ACS Nanoscience Au},
  volume = {3},
  number = {4},
  pages = {269--279},
  issn = {2694-2496, 2694-2496},
  doi = {10.1021/acsnanoscienceau.2c00061},
  urldate = {2025-12-04},
  copyright = {https://creativecommons.org/licenses/by/4.0/},
  langid = {english}
}

@article{hourahine2020DFTB,
  title = {{{DFTB}}+, a Software Package for Efficient Approximate Density Functional Theory Based Atomistic Simulations},
  author = {Hourahine, Ben and Aradi, B{\'a}lint and Blum, Volker and Bonafe, Frank and Buccheri, Alex and Camacho, Cristopher and Cevallos, Caterina and Deshaye, Megan Y and Dumitric{\u a}, Traian and Dominguez, Adriel and others},
  year = 2020,
  journal = {The Journal of chemical physics},
  volume = {152},
  number = {12},
  publisher = {AIP Publishing}
}

@article{hovel1993Width,
  title = {Width of Cluster Plasmon Resonances: {{Bulk}} Dielectric Functions and Chemical Interface Damping},
  shorttitle = {Width of Cluster Plasmon Resonances},
  author = {H{\"o}vel, H. and Fritz, S. and Hilger, A. and Kreibig, U. and Vollmer, M.},
  year = 1993,
  month = dec,
  journal = {Physical Review B},
  volume = {48},
  number = {24},
  pages = {18178--18188},
  issn = {0163-1829, 1095-3795},
  doi = {10.1103/PhysRevB.48.18178},
  urldate = {2025-12-04},
  copyright = {http://link.aps.org/licenses/aps-default-license},
  langid = {english}
}

@article{joao2023Atomistic,
  title = {Atomistic {{Theory}} of {{Hot-Carrier Relaxation}} in {{Large Plasmonic Nanoparticles}}},
  author = {Jo{\~a}o, Sim{\~a}o M. and Jin, Hanwen and Lischner, Johannes C.},
  year = 2023,
  month = dec,
  journal = {The Journal of Physical Chemistry C},
  volume = {127},
  number = {48},
  pages = {23296--23302},
  issn = {1932-7447, 1932-7455},
  doi = {10.1021/acs.jpcc.3c05347},
  urldate = {2025-12-04},
  copyright = {https://creativecommons.org/licenses/by/4.0/},
  langid = {english}
}

@book{kadanoff2000Quantum,
  title = {Quantum Statistical Mechanics: {{Green}}'s Function Methods in Equilibrium and Nonequilibrium Problems},
  shorttitle = {Quantum Statistical Mechanics},
  author = {Kadanoff, Leo P. and Baym, Gordon A.},
  year = 2000,
  series = {Advanced Book Classics Series},
  edition = {Nachdr.},
  publisher = {Perseus Books},
  address = {Cambridge, Mass},
  isbn = {978-0-201-09422-0 978-0-201-41046-4},
  langid = {english}
}

@article{kazuma2019Mechanistic,
  title = {Mechanistic {{Studies}} of {{Plasmon Chemistry}} on {{Metal Catalysts}}},
  author = {Kazuma, Emiko and Kim, Yousoo},
  year = 2019,
  month = apr,
  journal = {Angewandte Chemie International Edition},
  volume = {58},
  number = {15},
  pages = {4800--4808},
  issn = {1433-7851, 1521-3773},
  doi = {10.1002/anie.201811234},
  urldate = {2025-12-04},
  langid = {english}
}

@article{khurgin2019Hot,
  title = {Hot Carriers Generated by Plasmons: Where Are They Generated and Where Do They Go from There?},
  shorttitle = {Hot Carriers Generated by Plasmons},
  author = {Khurgin, Jacob B.},
  year = 2019,
  journal = {Faraday Discussions},
  volume = {214},
  pages = {35--58},
  issn = {1359-6640, 1364-5498},
  doi = {10.1039/C8FD00200B},
  urldate = {2025-12-05},
  langid = {english}
}

@article{kononov2022Electron,
  title = {Electron Dynamics in Extended Systems within Real-Time Time-Dependent Density-Functional Theory},
  author = {Kononov, Alina and Lee, Cheng-Wei and Dos Santos, Tatiane Pereira and Robinson, Brian and Yao, Yifan and Yao, Yi and Andrade, Xavier and Baczewski, Andrew David and Constantinescu, Emil and Correa, Alfredo A. and Kanai, Yosuke and Modine, Normand and Schleife, Andr{\'e}},
  year = 2022,
  month = sep,
  journal = {MRS Communications},
  volume = {12},
  number = {6},
  pages = {1002--1014},
  issn = {2159-6867},
  doi = {10.1557/s43579-022-00273-7},
  urldate = {2025-12-06},
  langid = {english}
}

@article{kotani2007Quasiparticlea,
  title = {Quasiparticle Self-Consistent {{G W}} Method: {{A}} Basis for the Independent-Particle Approximation},
  shorttitle = {Quasiparticle Self-Consistent {{G W}} Method},
  author = {Kotani, Takao and Van Schilfgaarde, Mark and Faleev, Sergey V.},
  year = 2007,
  month = oct,
  journal = {Physical Review B},
  volume = {76},
  number = {16},
  pages = {165106},
  issn = {1098-0121, 1550-235X},
  doi = {10.1103/PhysRevB.76.165106},
  urldate = {2025-12-23},
  copyright = {http://link.aps.org/licenses/aps-default-license},
  langid = {english}
}

@article{kuisma2015Localized,
  title = {Localized Surface Plasmon Resonance in Silver Nanoparticles: {{Atomistic}} First-Principles Time-Dependent Density-Functional Theory Calculations},
  shorttitle = {Localized Surface Plasmon Resonance in Silver Nanoparticles},
  author = {Kuisma, M. and Sakko, A. and Rossi, T. P. and Larsen, A. H. and Enkovaara, J. and Lehtovaara, L. and Rantala, T. T.},
  year = 2015,
  month = mar,
  journal = {Physical Review B},
  volume = {91},
  number = {11},
  pages = {115431},
  issn = {1098-0121, 1550-235X},
  doi = {10.1103/PhysRevB.91.115431},
  urldate = {2025-12-04},
  copyright = {http://link.aps.org/licenses/aps-default-license},
  langid = {english}
}

@article{ladstadter2004Firstprinciples,
  title = {First-Principles Calculation of Hot-Electron Scattering in Metals},
  author = {Ladst{\"a}dter, Florian and Hohenester, Ulrich and Puschnig, Peter and {Ambrosch-Draxl}, Claudia},
  year = 2004,
  month = dec,
  journal = {Physical Review B},
  volume = {70},
  number = {23},
  pages = {235125},
  issn = {1098-0121, 1550-235X},
  doi = {10.1103/PhysRevB.70.235125},
  urldate = {2025-12-04},
  copyright = {http://link.aps.org/licenses/aps-default-license},
  langid = {english}
}

@article{larsen2017atomic,
  title = {The Atomic Simulation Environment---a {{Python}} Library for Working with Atoms},
  author = {Larsen, Ask Hjorth and Mortensen, Jens J{\o}rgen and Blomqvist, Jakob and Castelli, Ivano E and Christensen, Rune and Du{\l}ak, Marcin and Friis, Jesper and Groves, Michael N and Hammer, Bj{\o}rk and Hargus, Cory and others},
  year = 2017,
  journal = {Journal of Physics: Condensed Matter},
  volume = {29},
  number = {27},
  pages = {273002},
  publisher = {IOP Publishing}
}

@article{lee2023Band,
  title = {{\emph{D}} -{{Band Hole Dynamics}} in {{Gold Nanoparticles Measured}} with {{Time-Resolved Emission Upconversion Microscopy}}},
  author = {Lee, Stephen A. and Kuhs, Christopher T. and Searles, Emily K. and Everitt, Henry O. and Landes, Christy F. and Link, Stephan},
  year = 2023,
  month = apr,
  journal = {Nano Letters},
  volume = {23},
  number = {8},
  pages = {3501--3506},
  issn = {1530-6984, 1530-6992},
  doi = {10.1021/acs.nanolett.3c00622},
  urldate = {2025-12-04},
  copyright = {https://doi.org/10.15223/policy-029},
  langid = {english}
}

@article{lerme2010Size,
  title = {Size {{Dependence}} of the {{Surface Plasmon Resonance Damping}} in {{Metal Nanospheres}}},
  author = {Lerm{\'e}, Jean and Baida, Hatim and Bonnet, Christophe and Broyer, Michel and Cottancin, Emmanuel and Crut, Aur{\'e}lien and Maioli, Paolo and Del Fatti, Natalia and Vall{\'e}e, Fabrice and Pellarin, Michel},
  year = 2010,
  month = oct,
  journal = {The Journal of Physical Chemistry Letters},
  volume = {1},
  number = {19},
  pages = {2922--2928},
  issn = {1948-7185, 1948-7185},
  doi = {10.1021/jz1009136},
  urldate = {2025-12-17},
  langid = {english}
}

@article{li2020RealTime,
  title = {Real-{{Time Time-Dependent Electronic Structure Theory}}},
  author = {Li, Xiaosong and Govind, Niranjan and Isborn, Christine and DePrince, A. Eugene and Lopata, Kenneth},
  year = 2020,
  month = sep,
  journal = {Chemical Reviews},
  volume = {120},
  number = {18},
  pages = {9951--9993},
  issn = {0009-2665, 1520-6890},
  doi = {10.1021/acs.chemrev.0c00223},
  urldate = {2025-12-04},
  copyright = {https://doi.org/10.15223/policy-029},
  langid = {english}
}

@article{linic2015Photochemical,
  title = {Photochemical Transformations on Plasmonic Metal Nanoparticles},
  author = {Linic, Suljo and Aslam, Umar and Boerigter, Calvin and Morabito, Matthew},
  year = 2015,
  journal = {Nature materials},
  volume = {14},
  number = {6},
  pages = {567--576},
  publisher = {Nature Publishing Group UK London}
}

@article{link1999Spectral,
  title = {Spectral {{Properties}} and {{Relaxation Dynamics}} of {{Surface Plasmon Electronic Oscillations}} in {{Gold}} and {{Silver Nanodots}} and {{Nanorods}}},
  author = {Link, Stephan and {El-Sayed}, Mostafa A.},
  year = 1999,
  month = oct,
  journal = {The Journal of Physical Chemistry B},
  volume = {103},
  number = {40},
  pages = {8410--8426},
  issn = {1520-6106, 1520-5207},
  doi = {10.1021/jp9917648},
  urldate = {2025-12-04},
  langid = {english}
}

@article{liu2018Relaxation,
  title = {Relaxation of {{Plasmon-Induced Hot Carriers}}},
  author = {Liu, Jun G. and Zhang, Hui and Link, Stephan and Nordlander, Peter},
  year = 2018,
  month = jul,
  journal = {ACS Photonics},
  volume = {5},
  number = {7},
  pages = {2584--2595},
  issn = {2330-4022, 2330-4022},
  doi = {10.1021/acsphotonics.7b00881},
  urldate = {2025-12-04},
  langid = {english}
}

@article{lugovskoy1999Ultrafast,
  title = {Ultrafast Electron Dynamics in Metals under Laser Irradiation},
  author = {Lugovskoy, Andrey V. and Bray, Igor},
  year = 1999,
  month = aug,
  journal = {Physical Review B},
  volume = {60},
  number = {5},
  pages = {3279--3288},
  issn = {0163-1829, 1095-3795},
  doi = {10.1103/PhysRevB.60.3279},
  urldate = {2025-12-04},
  copyright = {http://link.aps.org/licenses/aps-default-license},
  langid = {english}
}

@article{luo2023Plasmoninduced,
  title = {Plasmon-Induced Hot Carrier Dynamics and Utilization},
  author = {Luo, Jian and Wu, Qile and Zhou, Lin and Lu, Weixi and Yang, Wenxing and Zhu, Jia},
  year = 2023,
  journal = {Photonics Insights},
  volume = {2},
  number = {4},
  pages = {R08},
  issn = {2791-1748},
  doi = {10.3788/PI.2023.R08},
  urldate = {2025-12-04},
  langid = {english}
}

@article{lyu2023Photocatalysis,
  title = {Photocatalysis of {{Metallic Nanoparticles}}: {{Interband}} vs {{Intraband Induced Mechanisms}}},
  shorttitle = {Photocatalysis of {{Metallic Nanoparticles}}},
  author = {Lyu, Pin and Espinoza, Randy and Nguyen, Son C.},
  year = 2023,
  month = aug,
  journal = {The Journal of Physical Chemistry C},
  volume = {127},
  number = {32},
  pages = {15685--15698},
  issn = {1932-7447, 1932-7455},
  doi = {10.1021/acs.jpcc.3c04436},
  urldate = {2025-12-04},
  copyright = {https://creativecommons.org/licenses/by/4.0/},
  langid = {english}
}

@article{maitra2004Double,
  title = {Double Excitations within Time-Dependent Density Functional Theory Linear Response},
  author = {Maitra, Neepa T. and Zhang, Fan and Cave, Robert J. and Burke, Kieron},
  year = 2004,
  month = apr,
  journal = {The Journal of Chemical Physics},
  volume = {120},
  number = {13},
  pages = {5932--5937},
  issn = {0021-9606, 1089-7690},
  doi = {10.1063/1.1651060},
  urldate = {2025-12-04},
  langid = {english}
}

@article{manjavacas2014PlasmonInduced,
  title = {Plasmon-{{Induced Hot Carriers}} in {{Metallic Nanoparticles}}},
  author = {Manjavacas, Alejandro and Liu, Jun G. and Kulkarni, Vikram and Nordlander, Peter},
  year = 2014,
  month = aug,
  journal = {ACS Nano},
  volume = {8},
  number = {8},
  pages = {7630--7638},
  issn = {1936-0851, 1936-086X},
  doi = {10.1021/nn502445f},
  urldate = {2025-12-04},
  langid = {english}
}

@article{merschdorf2004Collective,
  title = {Collective and Single-Particle Dynamics in Time-Resolved Two-Photon Photoemission},
  author = {Merschdorf, M. and Kennerknecht, C. and Pfeiffer, W.},
  year = 2004,
  month = nov,
  journal = {Physical Review B},
  volume = {70},
  number = {19},
  pages = {193401},
  issn = {1098-0121, 1550-235X},
  doi = {10.1103/PhysRevB.70.193401},
  urldate = {2025-12-04},
  copyright = {http://link.aps.org/licenses/aps-default-license},
  langid = {english}
}

@article{mueller2013Relaxation,
  title = {Relaxation Dynamics in Laser-Excited Metals under Nonequilibrium Conditions},
  author = {Mueller, B. Y. and Rethfeld, B.},
  year = 2013,
  month = jan,
  journal = {Physical Review B},
  volume = {87},
  number = {3},
  pages = {035139},
  issn = {1098-0121, 1550-235X},
  doi = {10.1103/PhysRevB.87.035139},
  urldate = {2025-12-04},
  copyright = {http://link.aps.org/licenses/aps-default-license},
  langid = {english}
}

@article{pettine2023EnergyResolved,
  title = {Energy-{{Resolved Femtosecond Hot Electron Dynamics}} in {{Single Plasmonic Nanoparticles}}},
  author = {Pettine, Jacob and Maioli, Paolo and Vall{\'e}e, Fabrice and Del Fatti, Natalia and Nesbitt, David J.},
  year = 2023,
  month = jun,
  journal = {ACS Nano},
  volume = {17},
  number = {11},
  pages = {10721--10732},
  issn = {1936-0851, 1936-086X},
  doi = {10.1021/acsnano.3c02062},
  urldate = {2025-12-04},
  copyright = {https://doi.org/10.15223/policy-029},
  langid = {english}
}

@article{qian2005Lifetime,
  title = {Lifetime of a Quasiparticle in an Electron Liquid},
  author = {Qian, Zhixin and Vignale, Giovanni},
  year = 2005,
  month = feb,
  journal = {Physical Review B},
  volume = {71},
  number = {7},
  pages = {075112},
  issn = {1098-0121, 1550-235X},
  doi = {10.1103/PhysRevB.71.075112},
  urldate = {2025-12-04},
  copyright = {http://link.aps.org/licenses/aps-default-license},
  langid = {english}
}

@article{reinhard1996sum,
  title = {From Sum Rules to {{RPA}}: 3. {{Optical}} Dipole Response in Metal Clusters},
  shorttitle = {From Sum Rules to {{RPA}}},
  author = {Reinhard, P.-G. and Genzken, O. and Brack, M.},
  year = 1996,
  month = oct,
  journal = {Annalen der Physik},
  volume = {508},
  number = {7},
  pages = {576--607},
  issn = {0003-3804, 1521-3889},
  doi = {10.1002/andp.2065080704},
  urldate = {2025-12-04},
  langid = {english}
}

@article{rethfeld2002Ultrafast,
  title = {Ultrafast Dynamics of Nonequilibrium Electrons in Metals under Femtosecond Laser Irradiation},
  author = {Rethfeld, B. and Kaiser, A. and Vicanek, M. and Simon, G.},
  year = 2002,
  month = may,
  journal = {Physical Review B},
  volume = {65},
  number = {21},
  pages = {214303},
  issn = {0163-1829, 1095-3795},
  doi = {10.1103/PhysRevB.65.214303},
  urldate = {2025-12-04},
  copyright = {http://link.aps.org/licenses/aps-default-license},
  langid = {english}
}

@article{rossi2020HotCarrier,
  title = {Hot-{{Carrier Generation}} in {{Plasmonic Nanoparticles}}: {{The Importance}} of {{Atomic Structure}}},
  shorttitle = {Hot-{{Carrier Generation}} in {{Plasmonic Nanoparticles}}},
  author = {Rossi, Tuomas P. and Erhart, Paul and Kuisma, Mikael},
  year = 2020,
  month = aug,
  journal = {ACS Nano},
  volume = {14},
  number = {8},
  pages = {9963--9971},
  issn = {1936-0851, 1936-086X},
  doi = {10.1021/acsnano.0c03004},
  urldate = {2025-12-04},
  copyright = {http://pubs.acs.org/page/policy/authorchoice\_ccby\_termsofuse.html},
  langid = {english}
}

@article{runge1984Densityfunctional,
  title = {Density-Functional Theory for Time-Dependent Systems},
  author = {Runge, Erich and Gross, Eberhard KU},
  year = 1984,
  journal = {Physical Review Letters},
  volume = {52},
  number = {12},
  pages = {997},
  publisher = {APS}
}

@article{saavedra2016HotElectron,
  title = {Hot-{{Electron Dynamics}} and {{Thermalization}} in {{Small Metallic Nanoparticles}}},
  author = {Saavedra, J. R. M. and {Asenjo-Garcia}, Ana and Garc{\'i}a De Abajo, F. Javier},
  year = 2016,
  month = sep,
  journal = {ACS Photonics},
  volume = {3},
  number = {9},
  pages = {1637--1646},
  issn = {2330-4022, 2330-4022},
  doi = {10.1021/acsphotonics.6b00217},
  urldate = {2025-12-04},
  langid = {english}
}

@article{schlather2017Hot,
  title = {Hot {{Hole Photoelectrochemistry}} on {{Au}}@{{SiO}}{\textsubscript{2}} @{{Au Nanoparticles}}},
  author = {Schlather, Andrea E. and Manjavacas, Alejandro and Lauchner, Adam and Marangoni, Valeria S. and DeSantis, Christopher J. and Nordlander, Peter and Halas, Naomi J.},
  year = 2017,
  month = may,
  journal = {The Journal of Physical Chemistry Letters},
  volume = {8},
  number = {9},
  pages = {2060--2067},
  issn = {1948-7185, 1948-7185},
  doi = {10.1021/acs.jpclett.7b00563},
  urldate = {2026-01-21},
  langid = {english}
}

@article{seibel2023TimeResolved,
  title = {Time-{{Resolved Spectral Densities}} of {{Nonthermal Electrons}} in {{Gold}}},
  author = {Seibel, Christopher and Uehlein, Markus and Held, Tobias and Terekhin, Pavel N. and Weber, Sebastian T. and Rethfeld, Baerbel},
  year = 2023,
  month = dec,
  journal = {The Journal of Physical Chemistry C},
  volume = {127},
  number = {48},
  pages = {23349--23358},
  issn = {1932-7447, 1932-7455},
  doi = {10.1021/acs.jpcc.3c04581},
  urldate = {2025-12-04},
  copyright = {https://doi.org/10.15223/policy-029},
  langid = {english}
}

@article{sundararaman2014Theoretical,
  title = {Theoretical Predictions for Hot-Carrier Generation from Surface Plasmon Decay},
  author = {Sundararaman, Ravishankar and Narang, Prineha and Jermyn, Adam S. and Goddard Iii, William A. and Atwater, Harry A.},
  year = 2014,
  month = dec,
  journal = {Nature Communications},
  volume = {5},
  number = {1},
  pages = {5788},
  issn = {2041-1723},
  doi = {10.1038/ncomms6788},
  urldate = {2025-12-04},
  langid = {english}
}

@article{vanzan2024Theoretical,
  title = {Theoretical Approaches for the Description of Plasmon Generated Hot Carriers Phenomena},
  author = {Vanzan, Mirko and Marsili, Margherita},
  year = 2024,
  month = sep,
  journal = {npj Computational Materials},
  volume = {10},
  number = {1},
  pages = {222},
  issn = {2057-3960},
  doi = {10.1038/s41524-024-01412-5},
  urldate = {2025-12-04},
  langid = {english}
}

@article{varnavski2010Critical,
  title = {Critical {{Size}} for the {{Observation}} of {{Quantum Confinement}} in {{Optically Excited Gold Clusters}}},
  author = {Varnavski, Oleg and Ramakrishna, Guda and Kim, Junhyung and Lee, Dongil and Goodson, Theodore},
  year = 2010,
  month = jan,
  journal = {Journal of the American Chemical Society},
  volume = {132},
  number = {1},
  pages = {16--17},
  issn = {0002-7863, 1520-5126},
  doi = {10.1021/ja907984r},
  urldate = {2026-02-06},
  langid = {english}
}

@article{voisin2004Ultrafast,
  title = {Ultrafast Electron-Electron Scattering and Energy Exchanges in Noble-Metal Nanoparticles},
  author = {Voisin, C. and Christofilos, D. and Loukakos, P. A. and Del Fatti, N. and Vall{\'e}e, F. and Lerm{\'e}, J. and Gaudry, M. and Cottancin, E. and Pellarin, M. and Broyer, M.},
  year = 2004,
  month = may,
  journal = {Physical Review B},
  volume = {69},
  number = {19},
  pages = {195416},
  issn = {1098-0121, 1550-235X},
  doi = {10.1103/PhysRevB.69.195416},
  urldate = {2025-12-04},
  copyright = {http://link.aps.org/licenses/aps-default-license},
  langid = {english}
}

@article{wach2025dynamics,
  title = {The Dynamics of Plasmon-Induced Hot Carrier Creation in Colloidal Gold},
  author = {Wach, Anna and {Bericat-Vadell}, Robert and Bacellar, Camila and Cirelli, Claudio and Johnson, Philip J. M. and Castillo, Rebeca G. and Silveira, Vitor R. and Broqvist, Peter and Kullgren, Jolla and Maximenko, Alexey and Sobol, Tomasz and {Partyka-Jankowska}, Ewa and Nordlander, Peter and Halas, Naomi J. and Szlachetko, Jakub and S{\'a}, Jacinto},
  year = 2025,
  month = mar,
  journal = {Nature Communications},
  volume = {16},
  number = {1},
  pages = {2274},
  issn = {2041-1723},
  doi = {10.1038/s41467-025-57657-1},
  urldate = {2025-12-04},
  langid = {english}
}

@article{xu2024RealTime,
  title = {Real-{{Time Time-Dependent Density Functional Theory}} for {{Simulating Nonequilibrium Electron Dynamics}}},
  author = {Xu, Jianhang and Carney, Thomas E. and Zhou, Ruiyi and Shepard, Christopher and Kanai, Yosuke},
  year = 2024,
  month = feb,
  journal = {Journal of the American Chemical Society},
  volume = {146},
  number = {8},
  pages = {5011--5029},
  issn = {0002-7863, 1520-5126},
  doi = {10.1021/jacs.3c08226},
  urldate = {2025-12-04},
  copyright = {https://doi.org/10.15223/policy-029},
  langid = {english}
}

@article{yabana1996Timedependent,
  title = {Time-Dependent Local-Density Approximation in Real Time},
  author = {Yabana, Kazuhiro and Bertsch, {\relax GF}},
  year = 1996,
  journal = {Physical Review B},
  volume = {54},
  number = {7},
  pages = {4484},
  publisher = {APS}
}

@article{yannouleas1989Fragmentation,
  title = {Fragmentation of the Photoabsorption Strength in Neutral and Charged Metal Microclusters},
  author = {Yannouleas, C. and Broglia, R. A. and Brack, M. and Bortignon, P. F.},
  year = 1989,
  month = jul,
  journal = {Physical Review Letters},
  volume = {63},
  number = {3},
  pages = {255--258},
  issn = {0031-9007},
  doi = {10.1103/PhysRevLett.63.255},
  urldate = {2025-12-13},
  copyright = {http://link.aps.org/licenses/aps-default-license},
  langid = {english}
}

@article{yannouleas1992Landau,
  title = {Landau Damping and Wall Dissipation in Large Metal Clusters},
  author = {Yannouleas, C and Broglia, {\relax RA}},
  year = 1992,
  journal = {Annals of physics},
  volume = {217},
  number = {1},
  pages = {105--141},
  publisher = {Elsevier}
}

@article{zhang2018Plasmonic,
  title = {Plasmonic {{Hot-Carrier-Mediated Tunable Photochemical Reactions}}},
  author = {Zhang, Yu and Nelson, Tammie and Tretiak, Sergei and Guo, Hua and Schatz, George C.},
  year = 2018,
  month = aug,
  journal = {ACS Nano},
  volume = {12},
  number = {8},
  pages = {8415--8422},
  issn = {1936-0851, 1936-086X},
  doi = {10.1021/acsnano.8b03830},
  urldate = {2025-12-04},
  langid = {english}
}

@article{zhang2018SurfacePlasmonDriven,
  title = {Surface-{{Plasmon-Driven Hot Electron Photochemistry}}},
  author = {Zhang, Yuchao and He, Shuai and Guo, Wenxiao and Hu, Yue and Huang, Jiawei and Mulcahy, Justin R. and Wei, Wei David},
  year = 2018,
  month = mar,
  journal = {Chemical Reviews},
  volume = {118},
  number = {6},
  pages = {2927--2954},
  issn = {0009-2665, 1520-6890},
  doi = {10.1021/acs.chemrev.7b00430},
  urldate = {2025-12-04},
  langid = {english}
}

@article{zhang2019PlasmonDriven,
  title = {Plasmon-{{Driven Catalysis}} on {{Molecules}} and {{Nanomaterials}}},
  author = {Zhang, Zhenglong and Zhang, Chengyun and Zheng, Hairong and Xu, Hongxing},
  year = 2019,
  month = sep,
  journal = {Accounts of Chemical Research},
  volume = {52},
  number = {9},
  pages = {2506--2515},
  issn = {0001-4842, 1520-4898},
  doi = {10.1021/acs.accounts.9b00224},
  urldate = {2025-12-04},
  copyright = {https://doi.org/10.15223/policy-029},
  langid = {english}
}

@article{zhang2021Theory,
  title = {Theory of {{Plasmonic Hot-Carrier Generation}} and {{Relaxation}}},
  author = {Zhang, Yu},
  year = 2021,
  month = oct,
  journal = {The Journal of Physical Chemistry A},
  volume = {125},
  number = {41},
  pages = {9201--9208},
  issn = {1089-5639, 1520-5215},
  doi = {10.1021/acs.jpca.1c05837},
  urldate = {2025-12-04},
  copyright = {https://doi.org/10.15223/policy-029},
  langid = {english}
}

@article{zhao2024Relaxation,
  title = {Relaxation {{Channels}} of {{Two Types}} of {{Hot Carriers}} in {{Gold Nanostructures}}},
  author = {Zhao, Jie and Zhong, Yanyi and Zhang, Liyang and Sui, Laizhi and Wu, Guorong and Zhang, Jiangbin and Han, Kai and Zhang, Qi and Yuan, Kaijun and Yang, Xueming},
  year = 2024,
  month = dec,
  journal = {Nano Letters},
  volume = {24},
  number = {48},
  pages = {15340--15347},
  issn = {1530-6984, 1530-6992},
  doi = {10.1021/acs.nanolett.4c04431},
  urldate = {2025-12-04},
  copyright = {https://doi.org/10.15223/policy-029},
  langid = {english}
}

@article{zhou2017Evolution,
  title = {Evolution of {{Excited-State Dynamics}} in {{Periodic Au}}{\textsubscript{28}} , {{Au}}{\textsubscript{36}} , {{Au}}{\textsubscript{44}} , and {{Au}}{\textsubscript{52}} {{Nanoclusters}}},
  author = {Zhou, Meng and Zeng, Chenjie and Sfeir, Matthew Y. and Cotlet, Mircea and Iida, Kenji and Nobusada, Katsuyuki and Jin, Rongchao},
  year = 2017,
  month = sep,
  journal = {The Journal of Physical Chemistry Letters},
  volume = {8},
  number = {17},
  pages = {4023--4030},
  issn = {1948-7185, 1948-7185},
  doi = {10.1021/acs.jpclett.7b01597},
  urldate = {2026-02-06},
  langid = {english}
}

@article{zhukov2001Corrected,
  title = {Corrected Local-Density Approximation Band Structures, Linear-Response Dielectric Functions, and Quasiparticle Lifetimes in Noble Metals},
  author = {Zhukov, V. P. and Aryasetiawan, F. and Chulkov, E. V. and Gurtubay, I. G. De and Echenique, P. M.},
  year = 2001,
  month = oct,
  journal = {Physical Review B},
  volume = {64},
  number = {19},
  pages = {195122},
  issn = {0163-1829, 1095-3795},
  doi = {10.1103/PhysRevB.64.195122},
  urldate = {2025-12-04},
  copyright = {http://link.aps.org/licenses/aps-default-license},
  langid = {english}
}

\end{document}